\documentclass[onecolumn]{IEEEtran} 
\usepackage{setspace}
\usepackage{cite}
\usepackage{hyperref}
\ifCLASSINFOpdf
\usepackage[pdftex]{graphicx}
\else
   \usepackage[dvips]{graphicx}
\fi
\usepackage[cmex10]{amsmath}
\usepackage{algorithm}
\usepackage{algorithmic}
\usepackage{float}
\newfloat{algorithm}{t}{lop}
\usepackage{array}
\usepackage{booktabs}
\usepackage{url}
\usepackage{color}
\usepackage{multirow}
\usepackage{pdflscape}
\usepackage{afterpage}

\ifCLASSOPTIONcompsoc
  \usepackage[caption=false,font=normalsize,labelfont=sf,textfont=sf]{subfig}
\else
  \usepackage[caption=false,font=footnotesize]{subfig}
\fi
\usepackage{dblfloatfix}
\newcount\Comments  
\usepackage{color}
\definecolor{darkgreen}{rgb}{0,0.5,0}
\definecolor{purple}{rgb}{1,0,1}
\newcommand{\kibitz}[2]{\ifnum\Comments=0\textcolor{#1}{#2}\fi}

\begin{document}
\title{Model based string stability of adaptive cruise control systems using field data}

\author{George Gunter, Caroline Janssen, William Barbour, Raphael E. Stern, and Daniel~B.~Work,~\IEEEmembership{Member,~IEEE}
\thanks{G. Gunter is with the Department
of Civil and Environmental Engineering, University of Illinois at Urbana-Champaign, IL, and the Institute for Software Integrated Systms, Vanderbilt University, TN, email: (gunter1@illinois.edu).}
\thanks{C. Janssen and W. Barbour are with the Department of Civil and Environmental Engineering and Institute for Software Integrated Systems, Vanderbilt University, TN. email: (\{caroline.o.janssen, william.w.barbour\}@vanderbilt.edu).}
\thanks{R. Stern and D. Work are with the Department of Civil and Environmental Engineering and the Institute for Software Integrated Systems, Vanderbilt University, TN. email: (\{raphael.stern, dan.work\}@vanderbilt.edu).}
\thanks{*Corresponding author is R. Stern: raphael.stern@vanderbilt.edu}}



\maketitle
\begin{abstract}
This article is motivated by the lack of empirical data on the performance of commercially available Society of Automotive Engineers level one automated driving systems. To address this, a set of car following experiments are conducted to collect data from a 2015 luxury electric vehicle equipped with a commercial \textit{adaptive cruise control} (ACC) system. Velocity, relative velocity, and spacing data collected during the experiments are used to calibrate an optimal velocity relative velocity car following model for both the minimum and maximum following settings.  The string stability of both calibrated models is assessed, and it is determined that the best-fit models are string unstable, indicating they are not able to prevent all traffic disturbances from amplifying into phantom jams. Based on the calibrated models, we identify the consequences of the string unstable ACC system on synthetic and empirical lead vehicle disturbances, highlighting that some disturbances can be dampened even with string unstable commercial ACC platoons of moderate size.

\end{abstract}
\begin{IEEEkeywords}
Adaptive cruise control; phantom traffic jams; field experiments.
\end{IEEEkeywords}

 \ifCLASSOPTIONpeerreview
 \begin{center} \bfseries EDICS Category: 3-BBND \end{center}
\fi

\IEEEpeerreviewmaketitle

\section{Introduction}

Traffic jams that arise in the absence of bottlenecks are often referred to as \textit{phantom traffic jams}~\cite{orosz2009exciting, Helbing2001}. These may be stop-and-go waves where the vehicles come to a complete stop, or simply oscillatory traffic conditions that amplify as they propagate against the flow of traffic. While there are many common triggers that lead to phantom traffic jams, the seminal experiments of Sugiyama, et al.~\cite{Sugiyamaetal2008, Tadakietal2013} demonstrated that human driving behavior alone can be sufficient to trigger these waves. This finding was later verified by Wu, et al.~\cite{WuTRB2017, wu2018tracking}, who used a similar experimental setup and observed traffic waves emerging from human driving behavior, as well as Jiang, et al.~\cite{jiang2014traffic,jiang2017experimental}, who conducted a 51 vehicle platoon experiment and observed the emergence of phantom jams as a result of human driving behavior. These jams increase fuel consumption and emissions of the traffic flow~\cite{stern2017dissipation, stern2018emissions}.

To avoid phantom jams, it is important for a platoon of vehicles to be \textit{string stable}, meaning that small perturbations from an equilibrium flow are dissipated as they propagate from one vehicle to the next along the platoon~\cite{swaroop1996string}. The question of interest for phantom traffic jams is thus identifying whether a platoon of vehicles is string stable. This can be done by analyzing the car following dynamics of each vehicle in the platoon~\cite{wilson2011car}. Thus, assuming a homogeneous platoon where all vehicles follow the same dynamics, analyzing the behavior of a single vehicle pair is sufficient to identify the string stability of the overall vehicle platoon.

Interest in modeling vehicle dynamics at the individual vehicle level started in the 1950s when an expanding highway system promised to improve mobility, and it became clear that data was required to understand traffic at the level of the individual vehicle. Early and pioneering experimental efforts by researchers at General Motors collected velocity and spacing data to characterize driving behavior~\cite{rothery1964analysis, herman1959single, gazis1959car, chandler1958traffic}. The early experiments formed the basis of \textit{microscopic} traffic flow modelling, an area of study with numerous popular models that followed such as the \textit{Gipps model}~\cite{gipps1981behavioural}, the \textit{intelligent driver model}~\cite{treiber2000congested}, and the \textit{optimal velocity model} (OVM)~\cite{BandoHesebeNakayama1995}. Many of these models are able to reproduce the same type of instabilities seen in phantom jams~\cite{nakayama2009metastability,Cui2017}. 

One approach to prevent phantom jams from arising is to use connectivity and longitudinal vehicle control to form string stable vehicle platoons. Interest in platoons of string stable vehicles has existed for a while and it has been known that adding connectivity can guarantee stability and prevent phantom jams from arising within the platoon. This has been demonstrated both in theory~\cite{levine1966optimal, swaroop1996string,  darbha1999, besselink2017string,  buehler2009darpa} and experimentally~\cite{shladover1995review, fenton1991automated,ioannou1993intelligent,rajamani1998design}. 

More recently there has been interest in how a small number of \textit{autonomous vehicles} (AVs) are able to achieve string stability of a platoon even if not all vehicles in the flow are autonomous or have connectivity (e.g., mixed human and autonomous flows). This too has been considered both in theory~\cite{talebpour2016influence, davis2004effect,xiao2011practical,wang2018infrastructure} and experimentally~\cite{bose2001analysis,stern2017dissipation,jin2018experimental}. In the experiments conducted by Stern, et al.~\cite{stern2017dissipation}, a single autonomous vehicle in a stream of 20 human-piloted vehicles was able to stabilize the traffic flow and dampen stop-and-go waves. Recently, Jin, et al.~\cite{jin2018experimental} demonstrate experimentally that substantial improvements in fuel efficiency and safety may be achieved when only some vehicles use connected \textit{adaptive cruise control} (ACC).

Before vehicles become fully autonomous, it is likely that we will start to see an increasing number of \textit{Society of Automotive Engineers} (SAE) \textit{level one} and \textit{level two} automated vehicle systems on commercially available vehicles~\cite{bansal2017forecasting, litman2017autonomous}. When ACC controllers are designed with traffic stability in mind, they have been shown to have positive effects on the traffic flow~\cite{davis2004effect} and on vehicle emissions~\cite{bose2001analysis, ioannou2005evaluation}, even at a low market penetration rate. 

While ACC vehicles (without connectivity) have traditionally been considered a premium feature in luxury vehicles, more recently they have become a standard feature on many commercially available vehicles in the US. Through the second quarter of 2018, 16 of the 20 best selling cars in the US were available with ACC, and several of these vehicles were equipped with ACC as a standard feature~\cite{reuters}. This indicates the extent to which ACC vehicles are likely to become a common sight on US highways. Therefore, it is crucial to have a better understanding of the traffic stability implications of ACC vehicles that are now commercially available. 

The stability of ACC systems has been of interest for some time. In the early work by Bareket, et al.~\cite{bareket2003methodology}, a methodology is proposed and applied to by instrumenting vehicles with differential GPS receivers to collect relevant positioning and velocity data. 

After conducting a series of experiments on three ACC equipped vehicles in 2003, the work concluded, ``Based on measured characteristics of ACC systems, simulation analyzes [sic] indicate that currently-available ACC-equipped vehicles will have string-performance qualities that are characterized by substantial overshoots in velocity and range clearance in response to changes in the velocity of the preceding vehicle''~\cite{bareket2003methodology}. More recently in 2014,  Milan{\'e}s et al.~\cite{milanes2014modeling} instrumented a platoon of ACC vehicles and collected experimental data that also indicated the tested ACC system was string unstable.

One approach to assessing the string stability of an ACC system is outlined in~\cite{oncu2014cooperative} where the perturbation frequencies that are amplified from one vehicle to the next along the platoon of vehicles are identified. However, this approach requires the collection of data from a platoon of at least three vehicles to observe amplification in the spacing disturbance. Therefore, we consider an approach in which a model for ACC behavior is calibrated and then analyzed for string stability.

Our present work builds on the previous efforts to characterize the stability of ACC systems and addresses the question of whether some recent commercial ACC systems are unstable. Our main finding is that the tested commercial ACC system is string unstable, indicating that some disturbances will be amplified. We also show the consequences of string unstable ACC platoons on synthetic and empirical traffic disturbances.  

The string stability of the commercial ACC system is determined from a series of experimental car following tests. Using the collected data, a model of the ACC system is calibrated to and then used to determine the string stability of the vehicle and assess its consequences.  Given the sparsity of experimental work on the stability of commercially available ACC vehicles, this article provides important data and findings that will can help characterize the impacts of these systems on phantom traffic jams. We caution the reader that the results presented here do not indicate whether or not ACC vehicles perform better or worse than human drivers, which may also have string unstable dynamics~\cite{Sugiyamaetal2008}.

The remainder of the article is outlined as follows. In Section~\ref{sec:DynamicalModelAndStability} we review a common car following model that can be used to describe the dynamics of ACC equipped vehicles and compute the parameter regimes under which the model is string stable. In Section~\ref{sec:ExperimentalDescription}, an overview of the experimental setup, including vehicle instrumentation, and description of the testing procedure is provided. The methods used to estimate the model parameters from the data collected during the experiments are given in Section~\ref{sec:DataFittingMethodology}. In Section~\ref{sec:results}, the main results are presented indicating that under the best fit parameters, the ACC system of a recent, electric luxury sedan is string unstable, building on the findings reported on earlier commercially available ACC systems in 2014~\cite{milanes2014modeling}. We further illustrate the practical consequences of the system on realistic traffic disturbances.

\section{ACC Dynamical Model And Stability Analysis}\label{sec:DynamicalModelAndStability}
In this section modeling and analysis techniques are introduced that allow for the simulation and stability analysis of ACC-equipped vehicles. We first review an optimal velocity relative velocity car following model, and determine the parameter regimes under which the ACC model is string stable and string unstable. A brief numerical example shows the impact of the stability on the behaviour of a platoon of vehicles with ACC engaged.

\subsection{ACC model}
In general, high fidelity vehicle dynamics coupled with adaptive cruise control systems can be complex and difficult to replicate in simulation. The controllers may be implemented with logic determined by the vehicle state and environment~\cite{bareket2003methodology}, and depend on factors such as the engine RPM, the engine temperature, and the road grade. As such, approaches to completely replicate the exact control logic on commercial vehicle systems may be very difficult without good information about the internal vehicle state. Moreover, it may not be necessary to characterize the overall impacts of the ACC system on traffic flow stability. Consequently, we employ a car following model of an ACC vehicle, which models the vehicle dynamics and ACC as a single system. The model shows good performance when reconstructing the observed behavior of the ACC systems in field tests. The benefits of this simple model are that is easy to analyze and can readily be calibrated to field data.

Specifically, the adaptive cruise control is considered to be a behavioral rule that governs the acceleration $\dot{v}(t)$ of the following vehicle and is of the general form:
\begin{equation}
\label{eq:generic}
\begin{array}{rl}
\dot{s}(t) & =\Delta v\\
\dot{v}(t) &= f(s,v,\Delta v),
\end{array}
\end{equation}
where $v(t)$ is the velocity of the follower, $\Delta v:=v_l-v$ is the difference between the velocity of the leader (denoted $v_l$) and velocity of the follower, and $s$ is the \textit{space-gap}. The space-gap is defined as the distance between the rear bumper of the lead vehicle and the front bumper of the follower vehicle. It differs from the spacing (front bumper of the leader to front bumper of the follower) by the length of the lead vehicle.  Note that $\Delta v > 0$ indicates that the lead vehicle is going faster than the following vehicle.

One common car following model used to describe human driving dynamics is the \textit{optimal velocity} (OV) model~\cite{BandoHesebeNakayama1995}. The model takes the form:
\begin{equation}
\label{eq:OV}
    \dot{v}(t) = \alpha \left(V(s)-v\right).
\end{equation}
The OV model above represents a relaxation of the follower velocity to a desired velocity prescribed by the optimal velocity function $V$ based on the current spacing to the vehicle in front, and $\alpha$ is a model parameter.

A possible extension to the OV model is to add a term that relaxes the follower velocity to the velocity of the leader. This results in an OV model with a \textit{relative velocity} term (OVRV) and takes the form:

\begin{equation}
\label{eq:OVRV}
   \dot{v}(t) = \alpha \left(V(s)-v\right)+\beta \left(\Delta v \right).
\end{equation}
In the OVRV, the parameters $\alpha$ and $\beta$ control the trade-offs between following the optimal velocity and following the leader velocity.

For the purposes of modeling adaptive cruise control vehicles, we adopt the OVRV model with a special case of the OV component~\eqref{eq:OV} corresponding to a constant effective time-gap term~\cite{davis2013effects,liang1999optimal,milanes2014cooperative}:

\begin{equation}
\label{eq:OVRV-CTH}
\dot{v} = f(s,v,\Delta v) = k_1(s-\eta - \tau_e v)+k_2(\Delta v),
\end{equation}
where $\eta$ is the jam distance (space-gap when vehicles are completely stopped), the parameter $\tau_e$ is the desired effective time-gap, $k_1$ is the gain parameter on the constant effective time-gap term, $k_2$ is the gain parameter on the relative velocity term. Note that the model \eqref{eq:OVRV-CTH} operates under a linear optimal velocity function $V(s):=(s-\eta) /\tau_e$ and with $\alpha:=k_1\tau_e$. It is considered a constant effective time-gap term because the space-gap and velocity are adjusted based on the velocity such that the effective time-gap  $\tau_e$ is maintained.  It is well known that constant time-gap based controllers are important to overcome the inherent limitations of linear controllers to achieve a string stable constant spacing policy~\cite{seiler2004disturbance}. It is frequently used to model ACC systems because of its reported goodness of fit to simulate real trajectories of ACC equipped vehicles~\cite{milanes2014cooperative,liang1999optimal}.

We briefly note the importance of using an effective time-gap by explicitly including the jam distance term $\eta$ in~\eqref{eq:OVRV-CTH}, rather than the time-gap directly. Let $S(v):=v\tau(v)$ and consider an acceleration model $\dot v(t)=k(s-S(v))$, where $\tau(v)$ is the desired time-gap. In the special case where the time-gap $\tau(v)$ is a constant, the cars will collide at zero velocity (i.e., the car will continue to accelerate until $s=S(0)=0$). The following nonlinear time-gap model, 
\begin{equation}
\label{eq:nonlinear_time_gap}
\tau(v)=\eta/v+\tau_e,
\end{equation}

\noindent yields a spacing model $S(v)=\eta+\tau_e v$, and an acceleration model of $\dot v(t)=k(s-\eta -\tau_e v)$ which is precisely a constant effective time-gap model with effective time-gap of $\tau_e$. This turns out to be important when one fits empirical data collected from ACC vehicles in the sense that a nonlinear time-gap model \eqref{eq:nonlinear_time_gap} is in fact equivalent to a constant effective time-gap model.

\subsection{Stability analysis}
In this work the string stability of ACC enabled vehicles is examined. In broad terms, string stable driving behavior is critical to attenuate disturbances and prevent phantom jams from appearing from initially smooth and uniform flow. When a leading vehicle experiences a change in velocity in a string stable platoon, the following vehicles will experience a decreasing magnitude of response to the disturbance as it propagates through the platoon. In the other case where the platoon is string unstable, this perturbation will amplify as it propagates along the platoon.

We first assume the ACC vehicle dynamics satisfy the following \textit{rational driving constraints} (RDC)~\cite{wilson2011car}:

\begin{gather}
\label{eq:RDC}
\frac{\partial f}{\partial s} := f_{s} \geq 0, \\
\frac{\partial f}{\partial \Delta v} := f_{\Delta v} \geq 0, \\
\frac{\partial f}{\partial v} := f_{v} \leq 0.
\end{gather}

\noindent These intuitive conditions imply that as the the spacing or relative velocity increase, the follower should accelerate, and when the velocity decreases the follower will decelerate. 

The string stability of the following model of the ACC is considered:
\begin{equation}
\label{eq:modelwithd}
\begin{array}{rl}
\dot{s}(t) & =\Delta v\\
\dot{v}(t) &= k_1(s-\eta - \tau_e v)+k_2(\Delta v)+d,
\end{array}
\end{equation}
in which $d$ represents a disturbance to the acceleration. The corresponding velocity to velocity (also the space-gap to space-gap) transfer function reads~\cite{monteil2018l₂}:
\begin{equation}
\label{eq:transfer_function}
  \Gamma=\frac{zf_{\Delta v}+f_{s}}{z^{2}+z(f_{\Delta v}-f_{v})+f_{s}},
\end{equation}
where $z:=j\omega$ and $\omega \geq 0$ is the frequency. A sufficient condition for string stability of~\eqref{eq:modelwithd} is: 
\begin{equation}
\label{eq:stability_condition_on_transfer_function}
\left|\Gamma(j\omega)\right|=\sqrt{\frac{\omega^{2}f_{\Delta v}^{2}+f_{s}^{2}}{\left(f_{s}-\omega^{2}\right)^{2}+\omega^{2}\left(f_{\Delta v}-f_{v}\right)^{2}}}\leq 1,\quad \forall \omega \geq 0,
\end{equation}

\noindent see \cite{monteil2018l₂} for details. The condition~\eqref{eq:stability_condition_on_transfer_function} is equivalent to the well known conditions~\cite{wilson2011car}:

\begin{equation}
\label{eq:Stability Derivation Step 3}
\lambda_2:=\dfrac{f_{s}}{f_{v}^3}\left[\dfrac{f_{v}^2}{2} - f_{\Delta v}f_{v} - f_{s}\right]<0.
\end{equation}
To evaluate either condition, it suffices to compute the partial derivatives $f_s = k_1$,  $f_v = -k_1\tau_e$, and $f_{\Delta v} = k_2$, which depend only on the model parameters. Note that the jam-spacing $\eta$ does not affect the string stability of the model.

\begin{figure}
    \centering
    \includegraphics[trim=100 25 100 0,clip,width=0.6\columnwidth]{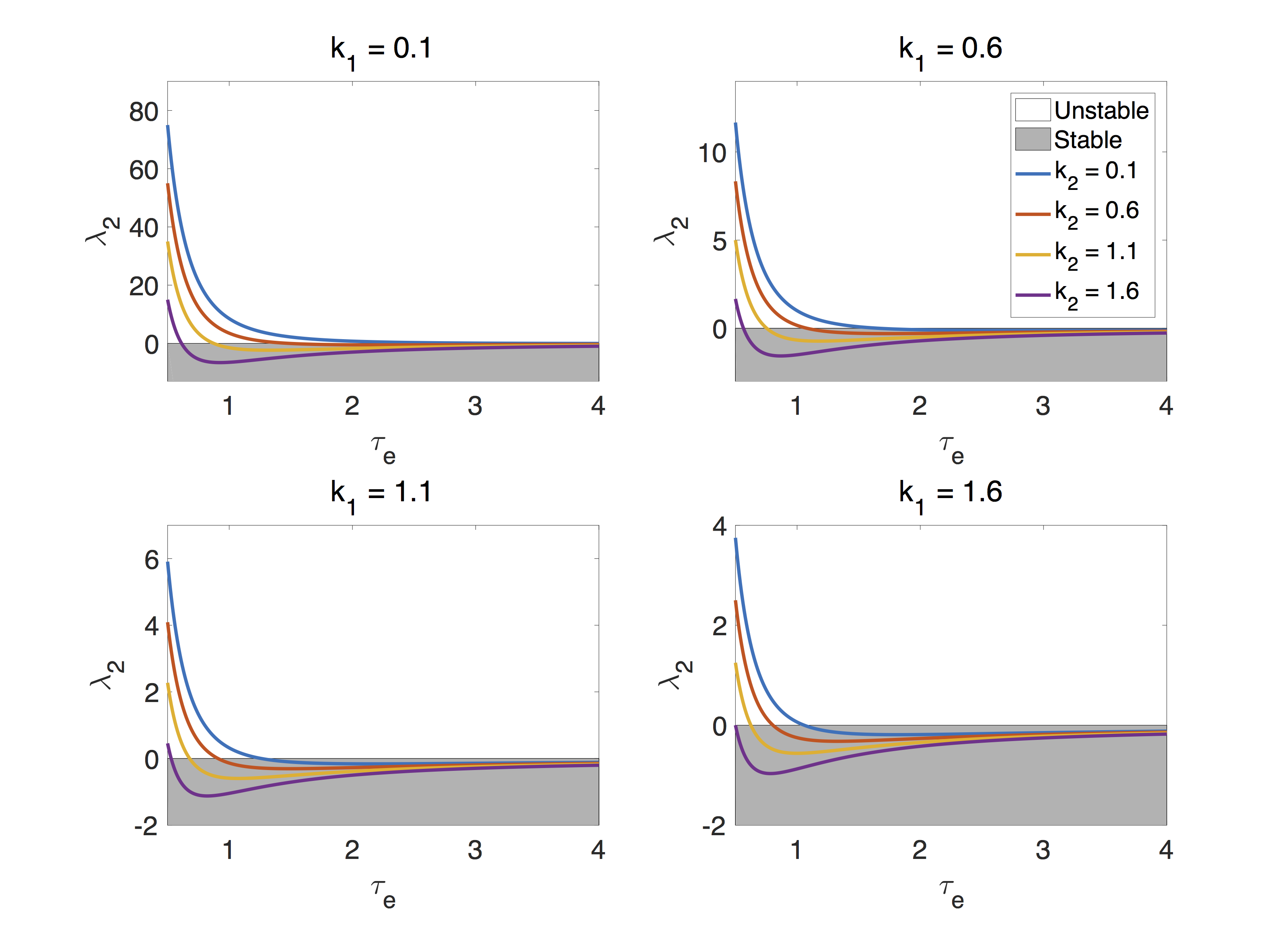}
    \caption{Stability criterion $\lambda_2$ for a range of gain values $k_1$ and $k_2$. The model is string stable for $\lambda_2<0$, indicated in grey.  For large $\tau_e$, the model is string stable but $\lambda_2$ approaches 0.}
    \label{fig:stability_plots}
\end{figure}

As an illustration in Figure~\ref{fig:stability_plots}, we determine the stability of~\eqref{eq:modelwithd} for ranges of $k_1$, $k_2$ and $\tau_e$. For models with small $k_1$ and $k_2$, a larger desired effective time-gap is necessary for string stability. Figure~\ref{fig:stability_plots} shows that increasing the effective time-gap $\tau_e$ may initially reduce $\lambda_2$, before eventually increasing $\lambda_2$. Note that for large $\tau_e$, the system is stable but $\lambda_2$ approaches 0. However, a consequence of a higher effective time-gap is that the traffic stream will have a lower throughput, since flow is inversely related to headway. 

\begin{figure}
\centering
\includegraphics[trim=50 0 100 0,clip,width=0.6\columnwidth]{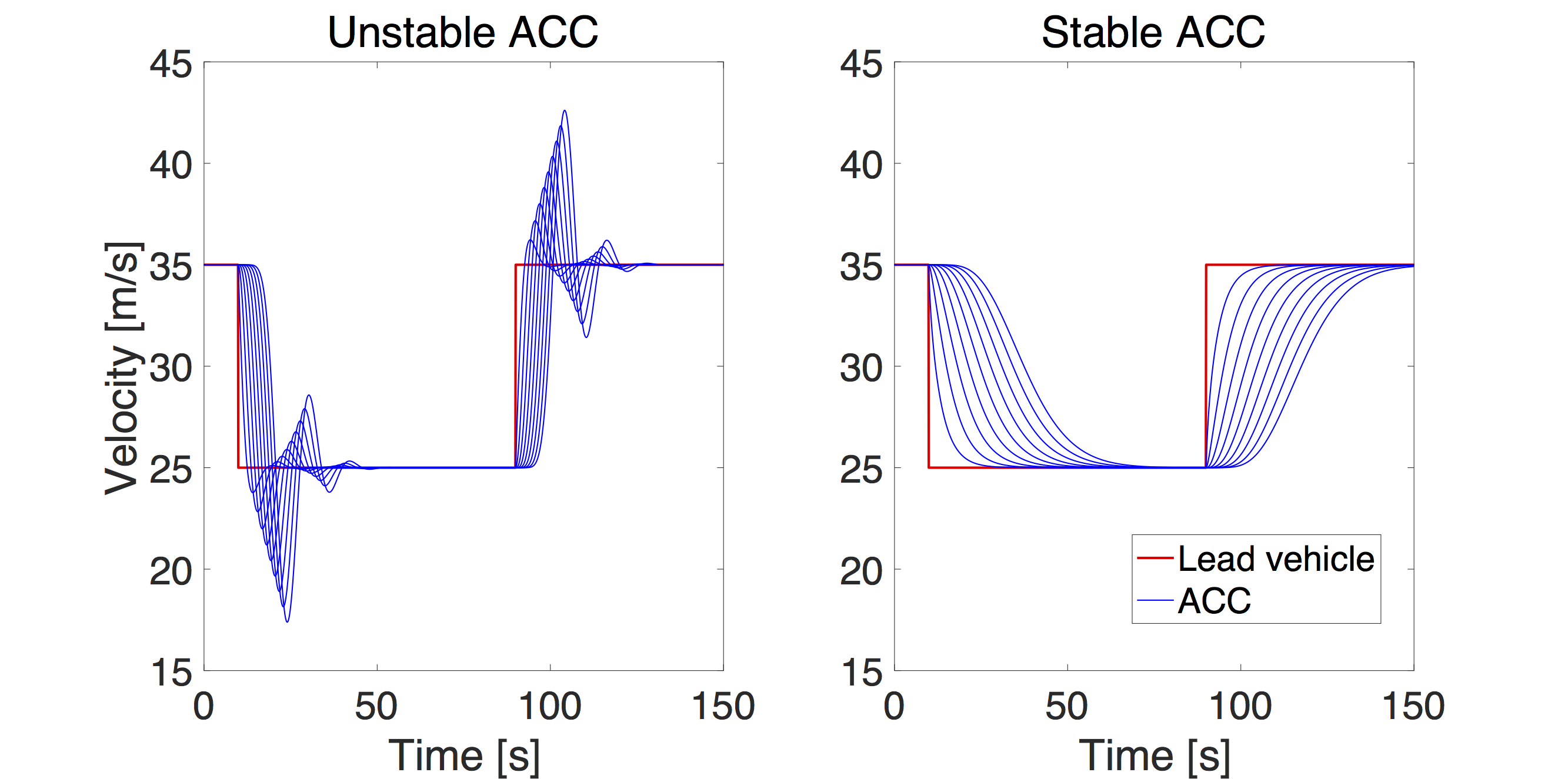}
\caption{Effect of varying $\tau_e$ on platoon string stability. The lead vehicle in red is followed by a platoon of nine ACC vehicles under the common parameters $k_1 = 0.5$, $k_2 = 0.5$, and $\eta = 8$. On the left, the effective time-gap $\tau_e = 0.75$ s results in a string unstable platoon. Under a larger time-gap of $\tau_e = 3.2$ s, the platoon on the right is string stable.}
\label{fig:wave stability comparison}
\end{figure}

An illustration of the consequences of string instability are provided in the form of a simulation where nine ACC equipped vehicles form a platoon behind a lead vehicle. All following vehicles proceed using the dynamical model in~\eqref{eq:OVRV-CTH}. The lead vehicle drives at a constant velocity then experiences a step-function decrease in velocity, and then after some time a following step-function increase back to the original velocity. In Figure \ref{fig:wave stability comparison}, each ACC equipped vehicle is simulated using~\eqref{eq:OVRV-CTH} with values of $k_1 = 0.5$, $k_2 = 0.5$, $\eta=8$, with the left figure using an effective time-gap $\tau_e = 0.75$ seconds and the right figure using $\tau_e = 3.2$ seconds. It is easy to verify that for $k_1 = 0.5$ and $k_2 = 0.5$ the two effective time-gaps represent respectively a string unstable system (left), and a string stable system (right). The left simulation displays significant overshoot both on the braking event and the acceleration event. The right simulation shows for the higher $\tau_e$ that the platoon does not overshoot either the braking or acceleration event and each following vehicle has a smoother response than the preceding vehicle.

\section{Experimental Overview and Test Vehicle Description}\label{sec:ExperimentalDescription}

\begin{figure}
    \centering
    \includegraphics[width = 0.9 \columnwidth]{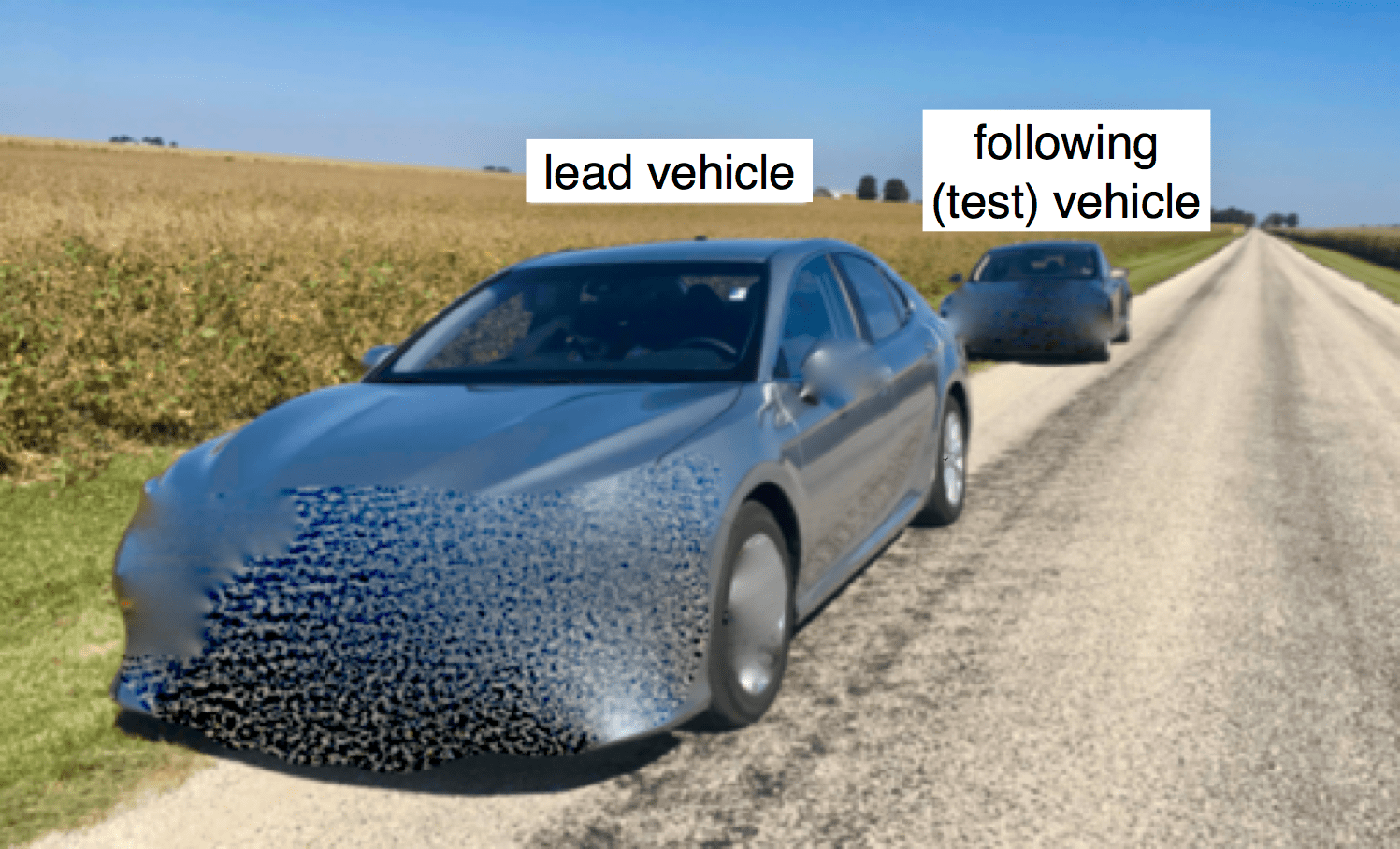}
    \caption{Vehicles used during experiment. Lead vehicle drives pre-specified velocity profile and following vehicle drives behind lead vehicle with ACC engaged.}
    \label{fig:experiment_setup}
\end{figure}

In this section we present the design and execution of a series of field experiments, with the goal to observe the following dynamics of an ACC-equipped following vehicle. Each experiment involves a lead vehicle that executes a pre-determined velocity profile and a following vehicle that follows the lead vehicle under adaptive cruise control  (Figure~\ref{fig:experiment_setup}).

The ACC system in the commercially-available vehicle tested in this experiment has two input settings: desired velocity and desired following setting 
(a minimum following setting, a maximum following setting, and several intermediate following settings). The desired velocity (to the nearest mile per hour), and the following setting are selected by the driver. 
In the tests conducted in this work, data is collected with the ACC engaged on either the closest or the furthest following setting, with the ACC desired velocity set at 5 mph above the maximum lead vehicle velocity for the given test. The ACC velocity setting ensures the vehicle remains in gap closing mode during the data collection. Tests are conducted on flat roadways with no hills or other topographic abnormalities. 

Each vehicle is equipped with a \textit{U-blox EVK-M8T} GPS evaluation kit that is capable of tracking the position and velocity of the vehicle throughout the experiment at a frequency of up to 10 Hz. Each evaluation kit is connected to a laptop computer, which runs a script to log the data as it is recorded. 

In each test, the vehicles are arranged with the lead vehicle in front under cruise control with the velocity selected by the driver, while the following (test) vehicle operates under control of the ACC system. A total of 18 tests are run using one of nine different lead vehicle velocity profiles and either the minimum or maximum following setting on the ACC follower vehicle.  In each test, the lead vehicle uses cruise control to execute the desired velocity profile for the test. In conditions where cruise control is not available on the lead vehicle  (i.e., for lead vehicle velocity profiles below 28 mph), the velocity profile is executed manually. The velocity profiles consist of a variety of steady car following conditions (i.e., where the lead vehicle follows a fixed velocity for a long period of time), and more dynamic conditions where the lead vehicle changes velocity quickly. The specific profiles (labeled A through I) are:

\begin{itemize}
    \item \textbf{velocity profile A:} \textit{low-velocity step test}, the lead vehicle travels at velocities ranging from 5 mph (2.2 m/s) to 30 mph (13.4 m/s) increasing the velocity by 5 mph at each step and holding the velocity for 60 seconds before moving to the next velocity. The same step function is followed again when decreasing the velocity from 30 mph to 5 mph.
    
    \item \textbf{velocity profile B:} \textit{medium-velocity step test}, the lead vehicle travels at velocities ranging from 35 mph (15.6 m/s) to 55 mph (24.6 m/s) increasing the velocity by 5 mph at each step and holding the velocity for at least 60 seconds before moving to the next velocity. The same step function is followed again when decreasing the velocity from 55 mph to 35 mph.
    
    \item \textbf{velocity profile C:} \textit{high-velocity step test}, the lead vehicle travels at velocities ranging from 60 mph (26.8 m/s) to 70 mph (31.3 m/s) increasing the velocity by 5 mph at each step and holding the velocity for 60 seconds before moving to the next velocity. The same step function is followed again when decreasing the velocity from 70 mph to 60 mph.
    
    \item \textbf{velocity profile D:} \textit{low-velocity oscillatory}, the lead vehicle oscillates between 30 mph and 20 mph (8.9 m/s) holding each velocity for 30 seconds.
    
    \item \textbf{velocity profile E:} \textit{medium-velocity 5 mph oscillatory}, the lead vehicle oscillates between 50 mph (22.4 m/s) and 45 mph (20.1 m/s) holding each velocity for 30 seconds.
    
    \item \textbf{velocity profile F:} \textit{medium-velocity 10 mph oscillatory}, the lead vehicle oscillates between 50 mph and 40 mph (17.9 m/s) holding each velocity for 30 seconds.
    
    \item \textbf{velocity profile G:} \textit{high-velocity 5 mph oscillatory}, the lead vehicle oscillates between 70 mph and 65 mph (29.1 m/s) holding each velocity for 30 seconds.
    
    \item \textbf{velocity profile H:} \textit{high-velocity 10 mph oscillatory}, the lead vehicle oscillates between 70 mph and 60 mph holding each velocity for 30 seconds.
    
    \item \textbf{velocity profile I:} \textit{medium-velocity dip}, the lead vehicle drives at 50 mph and conducts a series of rapid velocity decreases by 5 mph, 10 mph, 15 mph (6.7 m/s), and 20 mph holding each decreased velocity for 5 seconds and returning to 50 mph for 45 seconds after each velocity decrease.

\end{itemize}

\section{Model Calibration Methodology}\label{sec:DataFittingMethodology}

\subsection{Calibration of the ACC following dynamics}
In this section we outline how the ACC model~\eqref{eq:OVRV-CTH} is calibrated to the data collected during the driving experiments to determine the parameters  $k_1,k_2,\tau_e$, and $\eta$ that yield the best reconstruction of the observed data. The calibration is posed as a simulation-based optimization problem in which an error function is minimized by selecting optimal model parameters. In Milan\'es and Shladover~\cite{milanes2014modeling}, a mean absolute velocity error metric is considered, while in this work we instead consider an error metric based on the \textit{root mean square error} (RMSE) of the velocity:

\begin{equation}
    \label{eq:calibrationModel}
    \text{RMSE} = \sqrt{\frac{1}{T}\int_0^T{({v}_\text{m}(t)-v(t))^2dt}
}.
\end{equation}
In \eqref{eq:calibrationModel} the term $v(t)$ is the simulated velocity of the following vehicle at time $t$, ${v}_\text{m}(t)$ is the measured velocity of the following vehicle in the data at time $t$, and $T$ is the duration of the data collection period. Practically, in implementation, the RMSE is computed at the discrete time steps when measurements are collected (10 Hz).

The parameter values for each model (i.e., minimum and maximum following settings) are found using a constrained interior-point search method as implemented in the \texttt{fmincon} function in Matlab. The constraints consist of the initial spacing and velocity conditions, the assumed form of the dynamics of the adaptive cruise control system, and the rational driving constraints, which constrain the model to be physically realistic. For the ACC model, the rational driving constraints (and safety) imply $k_{1},k_{2},\tau_e$ (and $\eta$) are all non-negative. 
The simulation of the follower vehicle trajectory is solved using an explicit Euler step at 10 Hz (i.e., the sampling rate at which the data is collected). An explicit Runge-Kutta scheme~\cite{shampine1997matlab} was also considered, but was found to be slower while producing calibrated parameter values that were not substantially different from optimal parameters found using an explicit Euler step.

Summarizing, the parameter values $k_1k_2,\tau_e$ and $\eta$ are found by solving the following optimization problem constrained by the ACC dynamics, the initial conditions, and the rational driving constraints:

\begin{equation}\label{eq:minCTH}
\begin{array}{rl}
\underset{s,v,k_{1},k_{2},\tau_e,\eta}{{\text{minimize}}}: & \sqrt{\frac{1}{T}\int_0^T{({v}_\text{m}(t)-v(t))^2}dt}\\
\text{subject to:}
 & \dot{v}(t) = f(s, v, \Delta v)\\
 & \dot{s}(t) = v_{\ell,\text{m}}(t) - v(t)\\
 & s(0) = s_\text{m}(0)\\
 & v(0) = v_\text{m}(0)\\
 & k_{1} \geq 0\\
 & k_{2} \geq 0\\
 & \tau_e \geq 0\\
 & \eta \geq 0.
\end{array}
\end{equation}

\noindent In \eqref{eq:minCTH}, the term $v_{\ell,\text{m}}(t)$ denotes the measured velocity of the lead vehicle, which is used to evolve the spacing between the real lead vehicle and the simulated following vehicle. The initial space-gap $s(0)$ and the initial following velocity $v(0)$ are set as the initial measured spacing and initial measured following velocity of the ACC respectively.

The optimization problem \eqref{eq:minCTH} is nonlinear due to the fact that the state variables $s,v$ depend on the parameters to be calibrated, and consequently are also decision variables in the optimization problem. The nonlinear optimization problem potentially has local minima, so problem \eqref{eq:minCTH} is solved 100 times using randomly initialized parameters. The parameter values that yield the lowest RMSE out of the 100 runs are selected as the optimal parameter set for the model.

\section{Results}
\label{sec:results}
In this section we first provide an analysis of the accuracy of the GPS units that are used to measure vehicle positions and velocity. Next the calibration of the model \eqref{eq:OVRV-CTH} for both following settings is presented, and the results are compared to the measured ACC vehicle trajectory data. Finally, the stability of the calibrated dynamical models for the minimum and maximum following setting are determined and its consequences are described.

\subsection{Validation of GPS measurements}
The \textit{U-blox evaluation kit} GPS units are tested for relative velocity and positional accuracy by placing two U-blox sensors a known distance apart on the same vehicle and extensively driving this vehicle to observe the GPS measured distance and difference in velocity throughout the drive. 

\begin{figure}
    \centering
    \includegraphics[trim=50 0 100 0, clip, width=0.6\columnwidth]{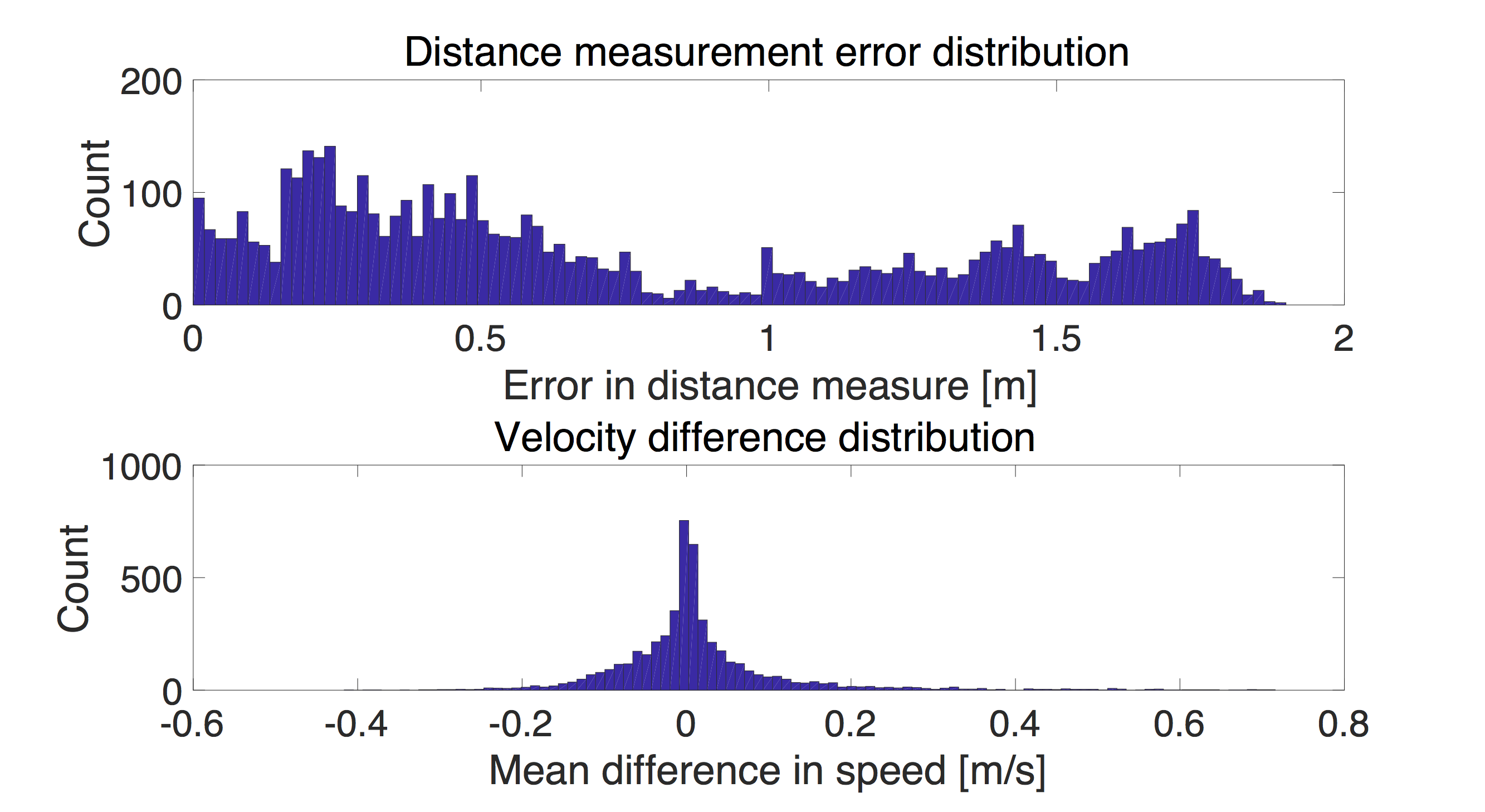}
    \caption{Distribution of error in position measurements between two U-blox EVK-M8T receivers mounted on the same vehicle (top) and distribution of instantaneous difference in velocity for same sensors (bottom).}
    \label{fig:vel_diff_hist}
\end{figure}

The distance between the two antennae mounted on the same vehicle is computed using the Haversine formula. The mean recorded sensor distance is 1.37~m while the actual sensor distance is 0.94~m. This represents a mean position accuracy accuracy of 0.43~m, which corresponds to 1--3\% error when compared to a typical following distance of between 15~m and 60~m, depending on the velocity. 
The mean absolute difference in velocity between the two sensors is 0.06~m/s (0.13~mph), which is an error of less than 3\% of the lowest velocity observed in the tests, with lower relative errors at higher velocity. 
The distribution of the relative position and velocity differences are shown in Figure~\ref{fig:vel_diff_hist}. Due to the overall good agreement between sensor velocity and position measurements, the U-blox EVK-M8T is a suitable GPS unit for recording position and velocity data.

\subsection{Model calibration and validation}\label{sec:ModelFittingVerification}

\afterpage{%
    \clearpage
    \thispagestyle{empty}
    \begin{landscape}
        \centering 
\begin{table*}
    \centering
        \begin{tabular}{ c c c c c c c c c c}
        \toprule
             velocity   & Following & Duration & Distance & Max velocity & Min velocity & velocity train & velocity test  & Space-gap       & Space-gap \\ 
              profile   & setting   & [s]      & [km]     & [km/h]       & [km/h]       & error [m/s]             & error [m/s]       & train error [m] & test error [m]\\
             \midrule
                
                A & minimum & 620 & 5.7  & 49.9  & 13.8 & 0.13 & 0.14 & 0.86 & 0.83\\
                B & minimum & 832 & 11.3 & 88.1  & 53.4 & 0.17 & 0.11 & 1.78 & 0.98\\
                C & minimum & 443 & 12.5 & 113.1 & 93.3 & 0.20 & 0.13 & 2.01 & 1.29\\
                D & minimum & 470 & 5.4  & 51.5  & 29.5 & 0.27 & 0.26 & 1.25 & 1.42\\
                E & minimum & 370 & 7.7  & 81.4  & 68.4 & 0.21 & 0.19 & 0.87 & 0.88\\
                F & minimum & 402 & 8.0  & 81.3  & 62.0 & 0.24 & 0.25 & 0.84 & 0.84\\
                G & minimum & 451 & 13.3 & 113.1 & 99.6 & 0.28 & 0.28 & 1.87 & 1.9\\
                H & minimum & 428 & 12.2 & 113.3 & 91.9 & 0.32 & 0.32 & 2.16 & 2.23\\
                I & minimum & 255 & 5.4  & 81.4  & 51.8 & 0.21 & 0.30 & 0.99 & 1.43\\
                \midrule
                Summary & minimum & -- & -- & -- & -- & 0.23
                & 0.22 & 1.51 & 1.37\\
                \bottomrule
                \\
                A & maximum & 551 & 4.8  & 49.2  & 16.3 & 0.13 & 0.16 & 2.87 & 2.2\\
                B & maximum & 744 & 11.2 & 89.8  & 53.1 & 0.22 & 0.11 & 2.99 & 1.28\\
                C & maximum & 458 & 12.9 & 113.5 & 92.7 & 0.27 & 0.13 & 2.84 & 1.56\\
                D & maximum & 409 & 4.7  & 49.4  & 31.0 & 0.32 & 0.30 & 2.66 & 2.71\\
                E & maximum & 383 & 7.9  & 81.9  & 68.4 & 0.23 & 0.26 & 2.48 & 2.48\\
                F & maximum & 391 & 7.8  & 81.7  & 62.2 & 0.21 & 0.21 & 3.27 & 3.2\\
                G & maximum & 496 & 14.6 & 113.7 & 98.8 & 0.40 & 0.44 & 3.59 & 3.97\\
                H & maximum & 498 & 14.2 & 113.7 & 91.9 & 0.41 & 0.42 & 3.42 & 3.55\\
                I & maximum & 307 & 6.6  & 81.7  & 53.7 & 0.26 & 0.37 & 2.56 & 2.66\\
                \midrule
                Summary & maximum & -- & -- & -- & -- & 0.28 & 0.30& 3.00 & 2.77\\
                
             \bottomrule
        \end{tabular}
        \vspace{0.1cm}
    \caption{Summary statistics for the following vehicle in each experiment.}
    \label{tab:Train&TestErrorForAllModels}
\end{table*}
    \end{landscape}
    \clearpage
}

\begin{table}
    \centering
        \begin{tabular}{c | c c c c c c c c c c c c}
        \toprule
             Following & $k_1$ & $k_2$ & $\tau_e$ & $\eta$ & $\lambda_2$ & String 
             \\
             ~ & [$1/\text{s} ^2$] & [1/s] & [s] & [m] & ~ & stability\\
             \midrule
                minimum & 0.0782 & 0.4445 & 0.5162 & 8.3365 & 70.7 & unstable\\ 
                maximum & 0.0131 & 0.2692 & 1.6881 & 7.5699 & 8.36 & unstable\\ 
             \bottomrule
        \end{tabular}
        \vspace{0.1cm}
    \caption{Calibrated model parameters and resulting string stability}
    \label{tab:CalibratedParametersAndLambda2}
\end{table}

\begin{figure}
\centering
\includegraphics[trim = 30 0 30 0, clip, width=0.6\columnwidth]{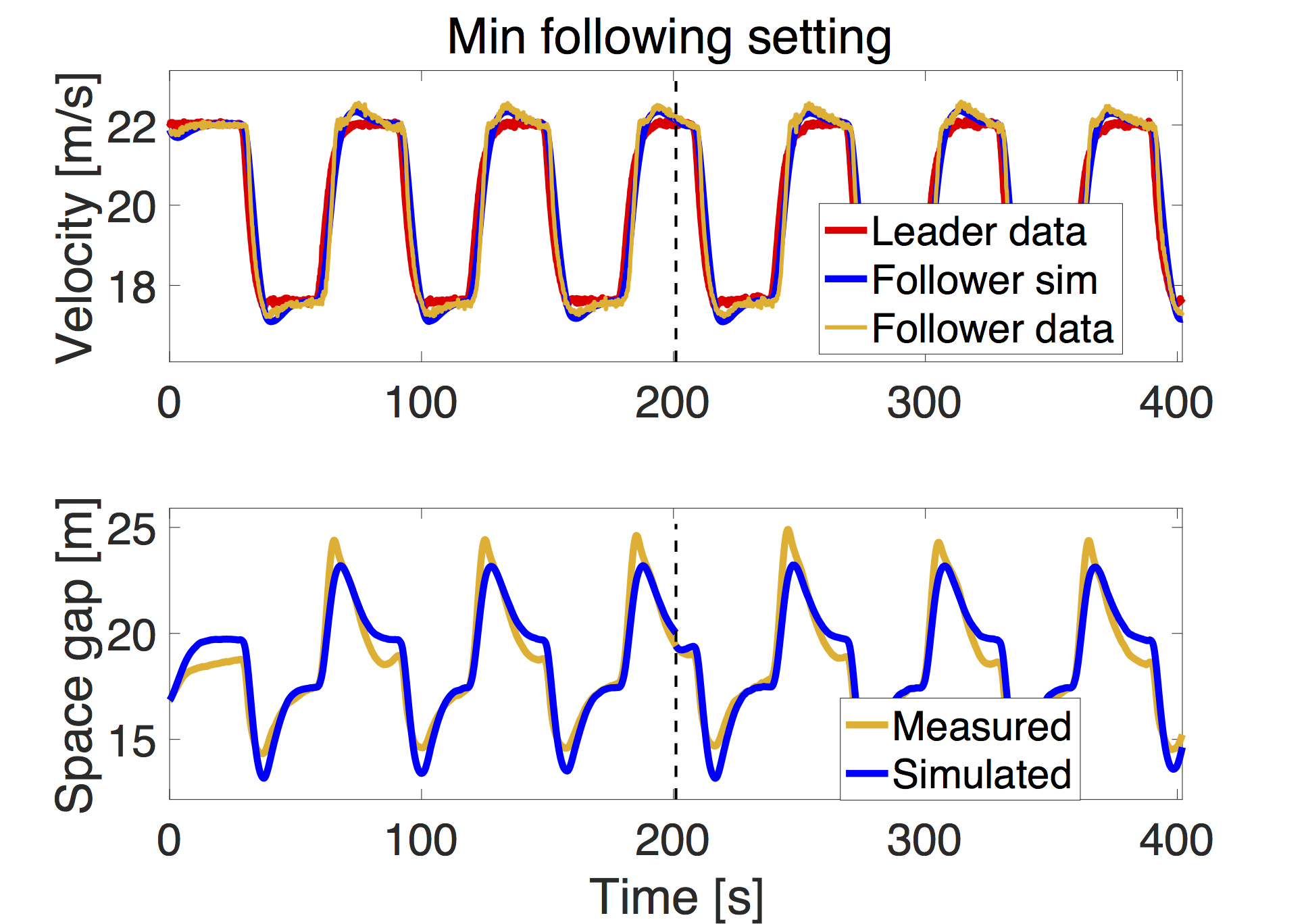}
\caption{Comparison between the empirical data and the calibrated ACC model under the minimum following setting for the velocity (top) and space-gap (bottom) for velocity profile F. The first 200 s is training data, while the remainder is the hold out test data.}
\label{fig:MinFollowingTrainTestTimeseries}
\end{figure}

\begin{figure}
\centering
\includegraphics[trim = 30 0 30 0, clip, width=0.6\columnwidth]{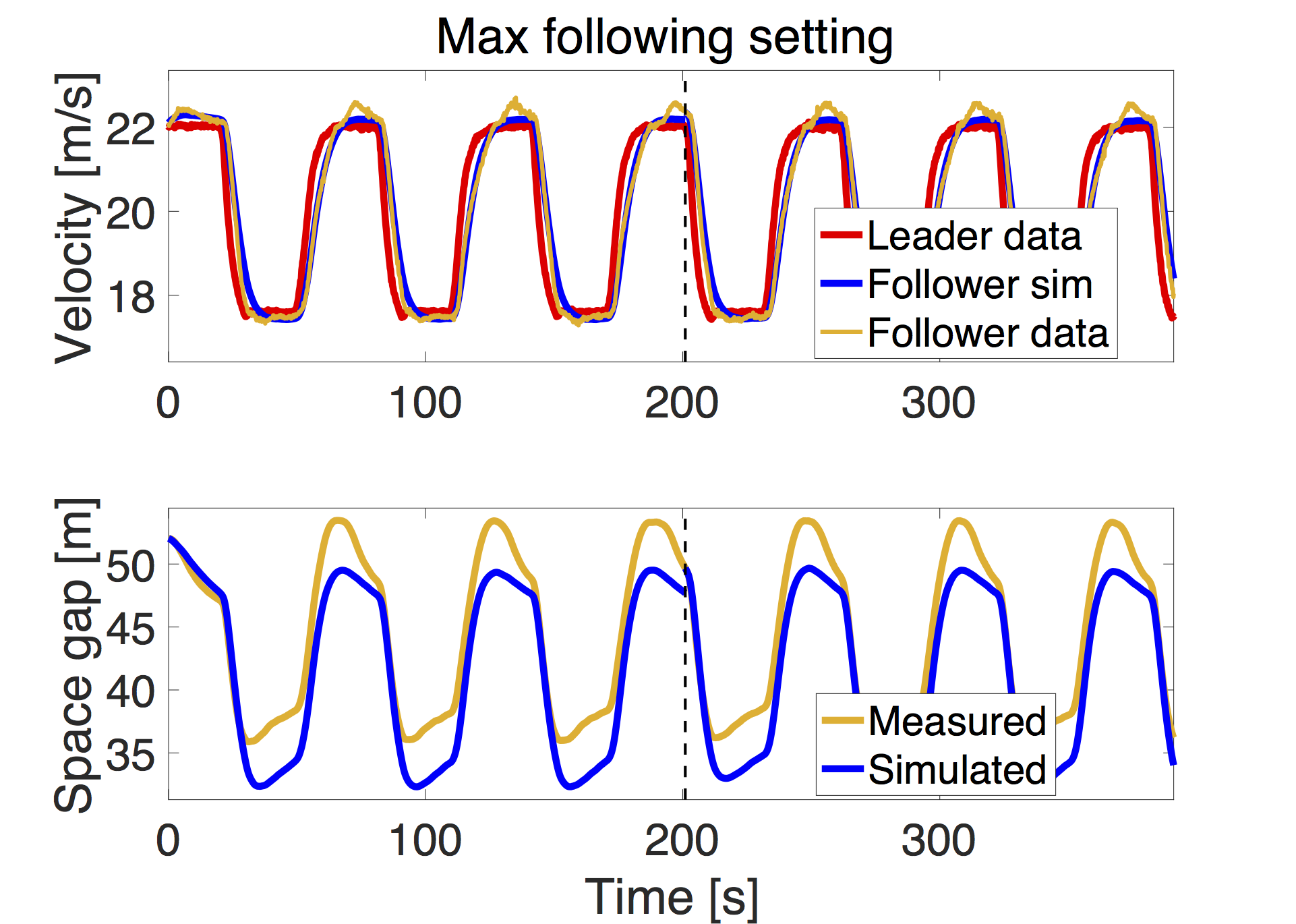}
\caption{Comparison between the empirical data and the calibrated ACC model under the maximum following setting for the velocity (top) and space-gap (bottom) for velocity profile F. The first 200 s is training data, while the remainder is the hold out test data.}
\label{fig:MaxFollowingTrainTestTimeseries}
\end{figure}

In this section, the calibration of the dynamical model in ~\eqref{eq:OVRV-CTH} to the experimental data collected is presented. This is done using the experiments from Section~\ref{sec:ExperimentalDescription} and the calibration routine outlined in Section \ref{sec:DataFittingMethodology}. 

For each following setting, we split each velocity profile in half. The first half of each velocity profile is used for training data, and the second half of each velocity profile is used as the hold out test set. A single model for each following setting is calibrated across all of the velocity profiles.  The overall training and test errors are reported as the summary values in Table~\ref{tab:Train&TestErrorForAllModels}, and the  best-fit calibrated model parameters are presented in Table~\ref{tab:CalibratedParametersAndLambda2}.  In addition to the RMSE velocity error~\eqref{eq:calibrationModel}, which is used as the performance measure to determine the best fitting parameters, we also report the space-gap RMSE errors.

For the minimum following setting, the RMSE training error across all velocity profiles is 0.23 m/s and 1.51 m for the velocity and space-gap, respectively. The test errors for the minimum following setting are 0.22 m/s and 1.37 m, which is slightly lower than the training error. The overall magnitude of the training and test errors are small and in good agreement, indicating the model is both a good fit, and is not overfitting the data. Exploring the performance of the model on the different velocity profiles, the lowest velocity test errors occur on the step tests (A, B, and C) that are near equilibrium, while the lowest space-gap errors occur on the medium velocity oscillatory tests (E and F). To help interpret the overall very good quality of fit,  in Figure~\ref{fig:MinFollowingTrainTestTimeseries} the velocity and space-gap are plotted for all of test F (medium velocity oscillatory), which has velocity (0.25 m/s) and space-gap (0.84 m) test errors. The figure shows that the calibrated model has good agreement with the observed data both for the training data (the first half of the test), and for the test data (the second half).

For the maximum following setting, a new model is calibrated and the overall quality of fit is slightly worse than the minimum following setting. The RMSE training errors are 0.28 m/s and 3.00 m for the velocity and space-gap, while the test errors are 0.30 m/s and 2.77 m. Again the training error and test errors are similar, indicating that the model is not overfitting the data. We again show the performance of the model on velocity profile E (medium velocity oscillations), which has velocity (0.26 m/s) and space-gap (2.48 m) test RMSE errors.

\begin{figure}
    \centering
    \includegraphics[trim = 70 0 70 0, clip, width=0.6\columnwidth]{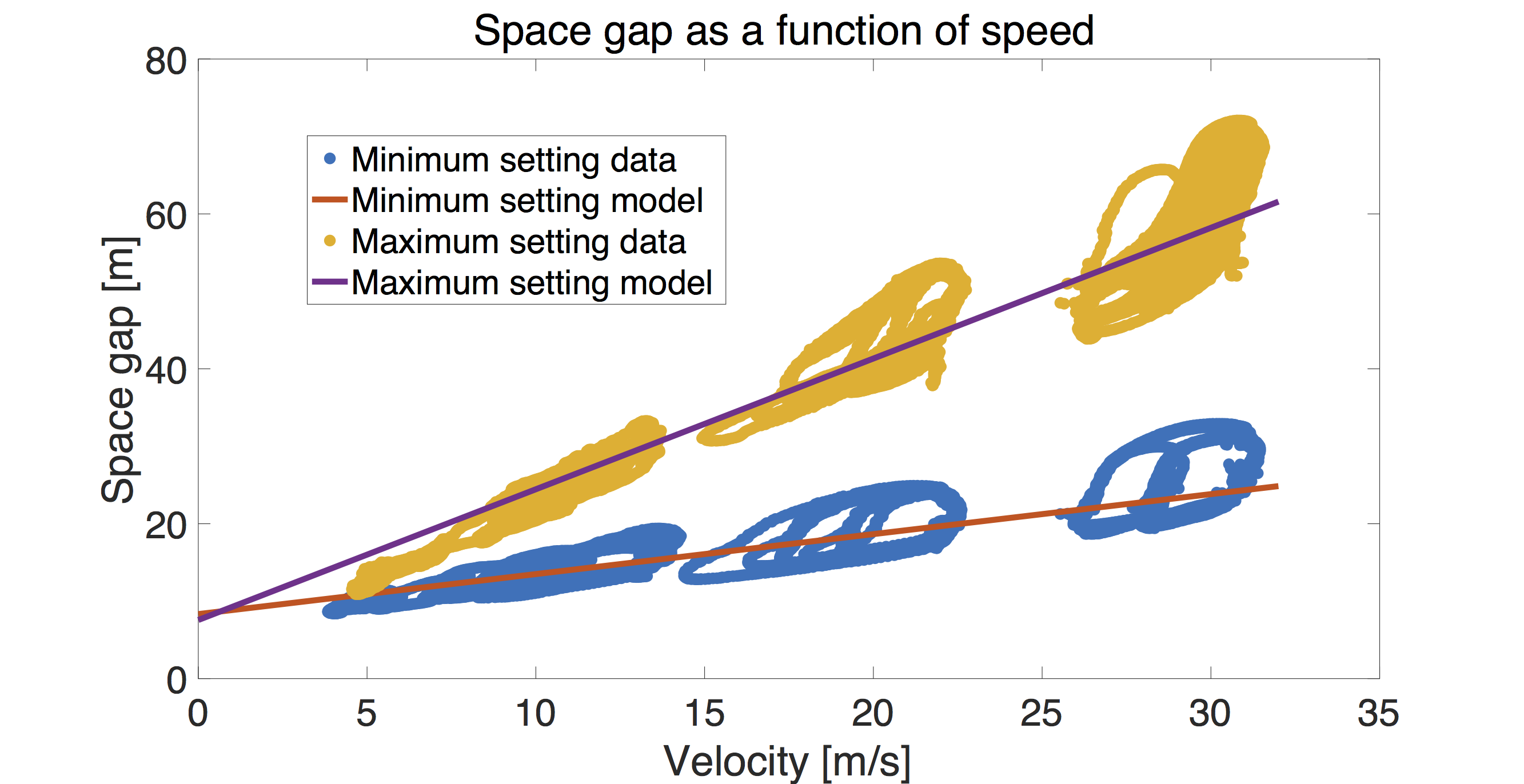}
    \caption{Space-gap as a function of the follower velocity for the calibrated model parameters and the empirical ACC data under both minimum and maximum following settings.}
    \label{fig:modelValidationFigure}
\end{figure}

The calibrated model parameters are also validated by comparing the velocity and space-gap observed in the data with the velocity and space-gap relationship that results from the calibrated model. This is presented in Figure~\ref{fig:modelValidationFigure}, where the relationship between velocity and space-gap resulting from the calibrated models for both the minimum and maximum following setting closely agree with the experimental data. The y-intercept corresponds to the jam distance $\eta$, which is in close agreement for both the minimum and maximum following settings. The difference in slopes corresponds to the different constant effective time-gaps (0.5 s for the minimum following setting, and 1.7 s for the maximum setting). 

\subsection{String stability of calibrated models}\label{sec:StabilityCheck}

\begin{figure}
    \centering
    \includegraphics[trim = 70 0 70 0, clip, width=0.6\columnwidth]{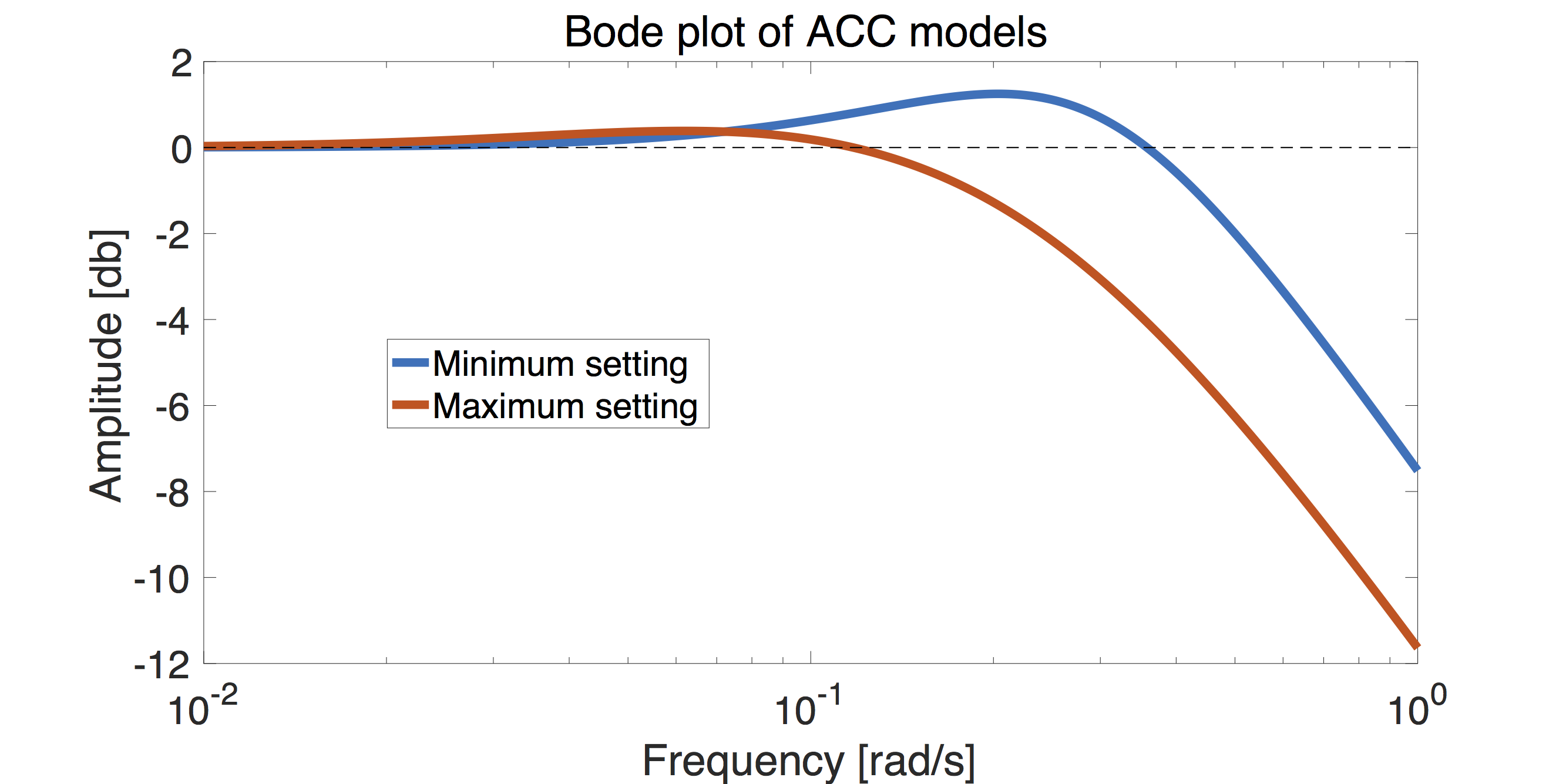}
    \caption{Bode plot of the calibrated ACC model under the minimum and maximum following settings.}
    \label{fig:BodePlot}
\end{figure}

In this section the stability of the calibrated models is calculated and discussed. First, the string stability criterion \eqref{eq:Stability Derivation Step 3} is calculated for each model. For the minimum following setting model under the calibrated parameters from~Table~\ref{tab:CalibratedParametersAndLambda2} $\lambda_2 = 70.7$ and for the maximum following setting model, $\lambda_2 = 8.36$.  Since $\lambda_2$ is non-negative for both the minimum and maximum following setting models, both settings are string unstable. This result indicates that under either following setting, perturbations to the traffic state may be amplified in magnitude as they propagate through  the platoon.

These results can be further explained by examining the Bode plots of the calibrated models (Figure~\ref{fig:BodePlot}). The Bode plot is generated for the velocity to velocity transfer function~\eqref{eq:transfer_function} evaluated using the calibrated model parameters. The amplitude of the transfer function in dB is given as a function of the frequency, where a positive amplitude indicates a disturbance at a given frequency will grow in magnitude as it propagates through the platoon, while a negative amplitude indicates that the disturbance will decay (see~\eqref{eq:stability_condition_on_transfer_function}). The ACC is string stable provided the amplitude of the transfer function is less than 0 dB for all frequencies. It can be seen from Figure~\ref{fig:BodePlot} that both models have portions of the frequency domain with a positive amplitude, and as such are string unstable.

While it is found from above analysis that the ACC system in consideration is string unstable under both following settings, there is a range of frequencies over which both ACC following settings will amplify disturbances, and also a range over which both ACC systems will dissipate them. For the minimum following setting, disturbances with frequencies less than 0.358 rad/s are amplified, while larger frequency disturbances will be dissipated along the platoon. Under the maximum following setting, the same is true for disturbances of frequency of 0.118 rad/s. The largest amplitude (1.25 dB) for the minimum following setting occurs at $\omega = 0.204$ rad/s, while the largest amplitude (0.386 dB) for the maximum following setting occurs at $\omega = 0.062$ rad/s.

\begin{figure}
    \centering
    \includegraphics[trim=80 0 80 0, clip, width=0.6\columnwidth]{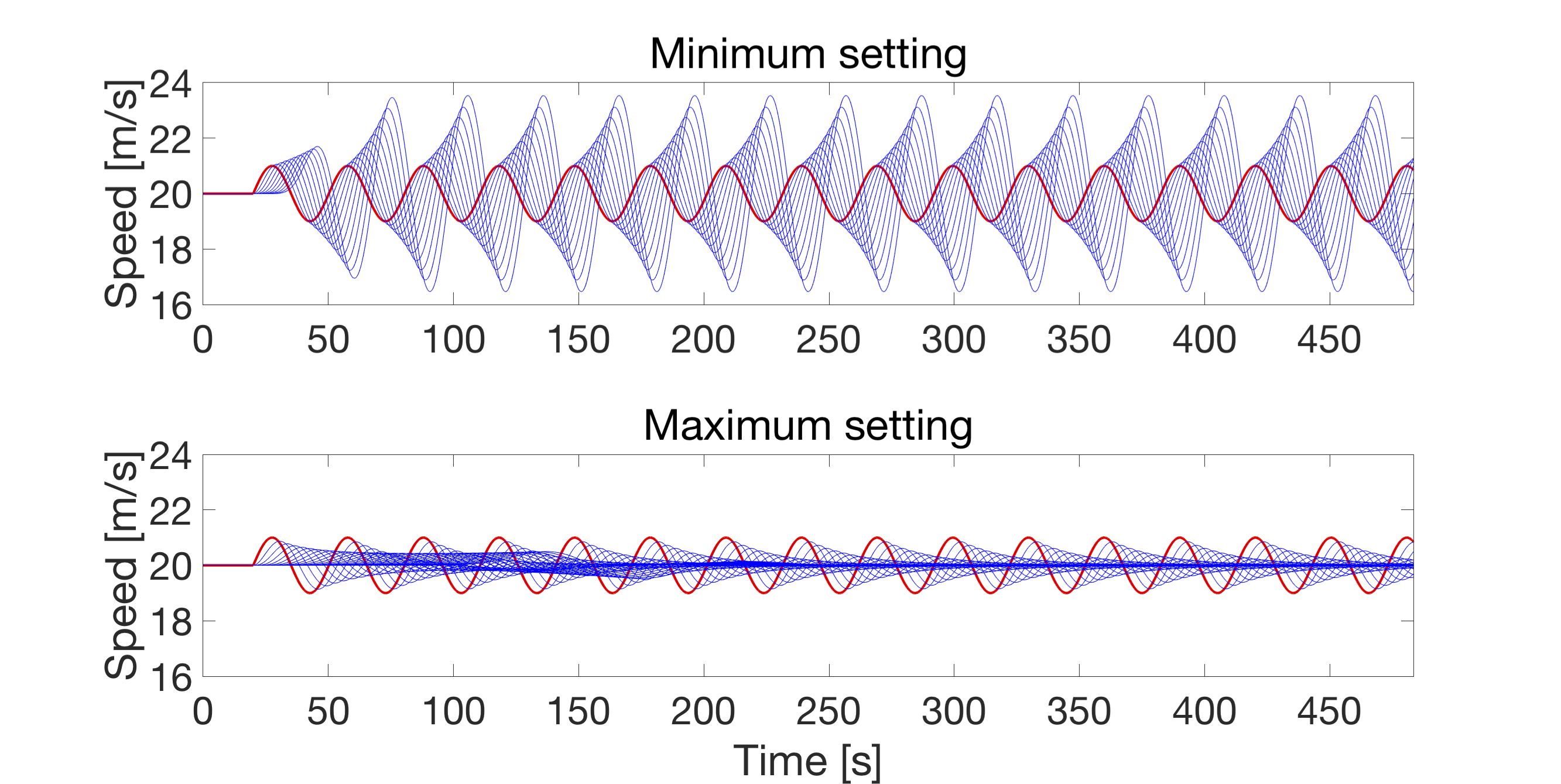}
    \caption{A simulation of 10 ACC vehicles for both minimum and maximum settings following a lead vehicle executing a sinusoidal velocity profile at $\omega = 0.204$ rad/s. The minimum following setting amplifies the perturbations, while the minimum dampens them.}
    \label{fig:PlatoonOscillationsComparison}
\end{figure}

To further illustrate that some frequencies are dissipated even with a string unstable ACC system, Figure~\ref{fig:PlatoonOscillationsComparison} shows the response of a 10 vehicle platoon to a lead vehicle executing a sinusoidal velocity pattern. The lead vehicle (shown in red) drives for 20 s with a velocity of 20 m/s with all following vehicles under ACC initialized at the corresponding equilibrium velocity and space-gap. After 20 s, the lead vehicle velocity follows a sinusoidal profile centered around the equilibrium velocity with a magnitude of 1 m/s and $\omega = 0.204$ rad/s, which is where the minimum following setting transfer function has the largest amplitude. The transfer function for the maximum following setting has a negative dB amplitude, meaning it will dissipate an oscillation of this frequency. As can be seen in Figure~\ref{fig:PlatoonOscillationsComparison}, the minimum following setting ACC amplifies the oscillation along the platoon, while the maximum setting ACC dissipates the disturbance.

\subsection{String unstable platoons following empirical lead vehicle velocity profiles}
\begin{figure}
    \centering
    \includegraphics[trim=40 0 40 0, clip, width=0.6\columnwidth]{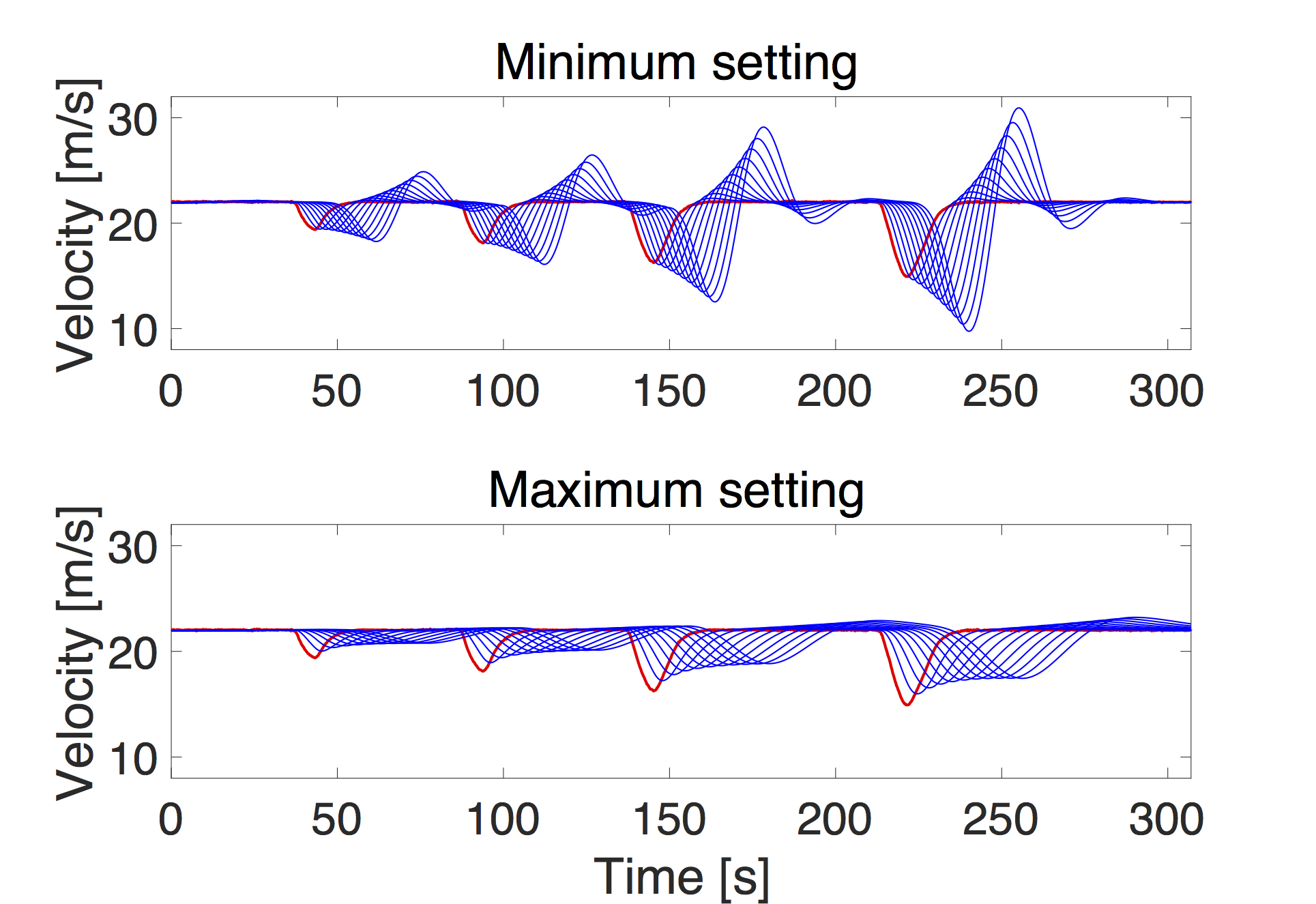}
    \caption{Consequences of lead vehicle disturbance following speed profile I for a 15 vehicle platoon for the minimum following setting and maximum following setting. The minimum following setting amplifies the disturbance while the minimum initially dampens them (before amplifying them for longer platoons).}
    \label{fig:RealDataPlatoon}
\end{figure}

In order to give a better understanding of the implication of the string instability of the calibrated models, a long ACC platoon is simulated following a lead vehicle driving according to the recorded velocity profile data from test I (medium-velocity dips). The lead vehicle velocity profile from test I represents a sudden slowdown by the lead vehicle and reflects a realistic braking event in the traffic flow that could trigger a phantom jam. Each vehicle in the ACC platoon is simulated using both the calibrated parameters under the minimum following setting, and the parameters for the maximum following setting. From Figure~\ref{fig:RealDataPlatoon}, we observe that the sudden braking event is amplified by the 15 vehicle ACC platoon under the minimum following setting, which is a consequence of the string unstable ACC. Interestingly, for the maximum following setting, the 15 vehicle ACC platoon initially dampens the disturbance, even though the ACC system is string unstable. However, for longer platoons (e.g., more than a 30 vehicle vehicle platoon), the velocity perturbation will eventually begin to grow again ultimately amplifying the initial disturbance. The consequence of the initial decay is that the overall magnitude of the disturbance is small for moderate sized platoons.

\section{Conclusions}
In this work car following experiments are conducted with a luxury electric sedan that is equipped with a commercially-available ACC system to collect data and fit a car following model that approximates the dynamics of the ACC system under minimum and maximum following settings.  The system is found to be string unstable, in line with the distinct commercial ACC system reported in Milan{\'e}s and Shladover in 2014~\cite{milanes2014modeling}. Consequently there are disturbance frequencies that are amplified as they propagate from one vehicle to another in a platoon. We also show that under practical disturbances such as a slow down event, the string unstable ACC systems are able to dampen disturbances for moderate sized platoons, even though they are eventually amplified for longer platoons.

\section{Acknowledgements}
This material is based upon work supported by the National Science Foundation under Grants No. CNS-1837652 (D.W. \& R.S.) and OISE-1743772 (G.G. \& C.J.) as well as through the Federal Highway Administration's Dwight David Eisenhower Transportation Fellowship Program under Grant No. 693JJ31845050 (R.S). The authors would also like to acknowledge Carl Gunter for providing the experimental vehicle as well as Derek Gloudemans, David Bergvelt, Hannah Burson, and Yanbing Wang for their assistance in data collection.

\bibliographystyle{IEEEtran}
\bibliography{refs}

\begin{thebibliography}{10}
\providecommand{\url}[1]{#1}
\csname url@samestyle\endcsname
\providecommand{\newblock}{\relax}
\providecommand{\bibinfo}[2]{#2}
\providecommand{\BIBentrySTDinterwordspacing}{\spaceskip=0pt\relax}
\providecommand{\BIBentryALTinterwordstretchfactor}{4}
\providecommand{\BIBentryALTinterwordspacing}{\spaceskip=\fontdimen2\font plus
\BIBentryALTinterwordstretchfactor\fontdimen3\font minus
  \fontdimen4\font\relax}
\providecommand{\BIBforeignlanguage}[2]{{%
\expandafter\ifx\csname l@#1\endcsname\relax
\typeout{** WARNING: IEEEtran.bst: No hyphenation pattern has been}%
\typeout{** loaded for the language `#1'. Using the pattern for}%
\typeout{** the default language instead.}%
\else
\language=\csname l@#1\endcsname
\fi
#2}}
\providecommand{\BIBdecl}{\relax}
\BIBdecl

\bibitem{orosz2009exciting}
G.~Orosz, R.~E. Wilson, R.~Szalai, and G.~St{\'e}p{\'a}n, ``Exciting traffic
  jams: nonlinear phenomena behind traffic jam formation on highways,''
  \emph{Physical Review E}, vol.~80, no.~4, p. 046205, 2009.

\bibitem{Helbing2001}
D.~Helbing, ``Traffic and related self-driven many-particle systems,''
  \emph{Reviews of Modern Physics}, vol.~73, pp. 1067--1141, Dec 2001.

\bibitem{Sugiyamaetal2008}
Y.~Sugiyama, M.~Fukui, M.~Kikuchi, K.~Hasebe, A.~Nakayama, K.~Nishinari,
  S.~i.~Tadaki, and S.~Yukawa, ``Traffic jams without bottlenecks --
  experimental evidence for the physical mechanism of the formation of a jam,''
  \emph{New Journal of Physics}, vol.~10, no.~3, p. 033001, 2008.

\bibitem{Tadakietal2013}
S.~Tadaki, M.~Kikuchi, M.~Fukui, A.~Nakayama, K.~Nishinari, A.~Shibata,
  Y.~Sugiyama, T.~Yosida, and S.~Yukawa, ``Phase transition in traffic jam
  experiment on a circuit,'' \emph{New Journal of Physics}, vol.~15, p. 103034,
  2013.

\bibitem{WuTRB2017}
F.~Wu, R.~E. Stern, M.~Churchill, M.~L.~D. Monache, K.~Han, B.~Piccoli, and
  D.~Work, ``Measuring trajectories and fuel consumption in oscillatory
  traffic: Experimental results,'' in \emph{Proceedings of the $96^{th}$
  Transportation Research Board Annual Meeting}, 2017.

\bibitem{wu2018tracking}
F.~Wu, R.~Stern, S.~Cui, M.~L. Delle~Monache, R.~Bhadani, M.~Bunting,
  M.~Churchill, N.~Hamilton, R.~Haulcy, H.~Pohlmann, B.~Piccoli, B.~Seibold,
  J.~Sprinkle, and D.~B. Work, ``Tracking vehicle trajectories and fuel rates
  in oscillatory traffic,'' \emph{Transportation Research Part C: Emerging
  Technologies, to appear}, 2018.

\bibitem{jiang2014traffic}
R.~Jiang, M.-B. Hu, H.~M. Zhang, Z.-Y. Gao, B.~Jia, Q.-S. Wu, B.~Wang, and
  M.~Yang, ``Traffic experiment reveals the nature of car-following,''
  \emph{PloS one}, vol.~9, no.~4, p. e94351, 2014.

\bibitem{jiang2017experimental}
R.~Jiang, C.-J. Jin, H.~M. Zhang, Y.-X. Huang, J.-F. Tian, W.~Wang, M.-B. Hu,
  H.~Wang, and B.~Jia, ``Experimental and empirical investigations of traffic
  flow instability,'' \emph{Transportation Research Part C: Emerging
  Technologies}, 2017.

\bibitem{stern2017dissipation}
R.~E. Stern, S.~Cui, M.~L.~D. Monache, R.~Bhadani, M.~Bunting, M.~Churchill,
  N.~Hamilton, R.~Haulcy, H.~Pohlmann, F.~Wu, B.~Piccoli, B.~Seibold,
  J.~Sprinkle, and D.~B. Work, ``Dissipation of stop-and-go waves via control
  of autonomous vehicles: Field experiments,'' \emph{Transportation Research
  Part C: Emerging Technologies}, vol.~89, pp. 205 -- 221, 2018.

\bibitem{stern2018emissions}
R.~E. Stern, Y.~Chen, M.~Churchill, F.~Wu, M.~L. Delle~Monache, B.~Piccoli,
  B.~Seibold, J.~Sprinkle, and D.~B. Work, ``Quantifying air quality benefits
  resulting from few autonomous vehicles stabilizing traffic,''
  \emph{Transportation Research Part D: Transport and Environment}, vol.~67,
  pp. 351--365, 2019.

\bibitem{swaroop1996string}
D.~Swaroop and J.~Hedrick, ``String stability of interconnected systems,''
  \emph{IEEE Transactions on Automatic Control}, vol.~41, no.~3, pp. 349--357,
  1996.

\bibitem{wilson2011car}
R.~E. Wilson and J.~A. Ward, ``Car-following models: fifty years of linear
  stability analysis--a mathematical perspective,'' \emph{Transportation
  Planning and Technology}, vol.~34, no.~1, pp. 3--18, 2011.

\bibitem{rothery1964analysis}
R.~Rothery, R.~Silver, R.~Herman, and C.~Torner, ``Analysis of experiments on
  single-lane bus flow,'' \emph{Operations Research}, vol.~12, no.~6, pp.
  913--933, 1964.

\bibitem{herman1959single}
R.~Herman and R.~B. Potts, ``Single lane traffic theory and experiment,'' in
  \emph{Proceedings of the Theory of Traffic Flow Symposium}, 1961, pp.
  120--146.

\bibitem{gazis1959car}
D.~C. Gazis, R.~Herman, and R.~B. Potts, ``Car-following theory of steady-state
  traffic flow,'' \emph{Operations Research}, vol.~7, no.~4, pp. 499--505,
  1959.

\bibitem{chandler1958traffic}
R.~E. Chandler, R.~Herman, and E.~W. Montroll, ``Traffic dynamics: studies in
  car following,'' \emph{Operations Research}, vol.~6, no.~2, pp. 165--184,
  1958.

\bibitem{gipps1981behavioural}
P.~G. Gipps, ``A behavioural car-following model for computer simulation,''
  \emph{Transportation Research Part B: Methodological}, vol.~15, no.~2, pp.
  105--111, 1981.

\bibitem{treiber2000congested}
M.~Treiber, A.~Hennecke, and D.~Helbing, ``Congested traffic states in
  empirical observations and microscopic simulations,'' \emph{Physical Review
  E}, vol.~62, no.~2, p. 1805, 2000.

\bibitem{BandoHesebeNakayama1995}
M.~Bando, H.~K., A.~Nakayama, A.~Shibata, and Y.~Sugiyama, ``Dynamical model of
  traffic congestion and numerical simulation,'' \emph{Physical Review E},
  vol.~51, no.~2, pp. 1035--1042, 1995.

\bibitem{nakayama2009metastability}
A.~Nakayama, M.~Fukui, M.~Kikuchi, K.~Hasebe, K.~Nishinari, Y.~Sugiyama, S.-i.
  Tadaki, and S.~Yukawa, ``Metastability in the formation of an experimental
  traffic jam,'' \emph{New Journal of Physics}, vol.~11, no.~8, p. 083025,
  2009.

\bibitem{Cui2017}
S.~Cui, B.~Seibold, R.~Stern, and D.~B. Work, ``Stabilizing traffic flow via a
  single autonomous vehicle: Possibilities and limitations,'' in
  \emph{Proceedings of the IEEE Intelligent Vehicles Symposium}, 2017, pp.
  1336--1341.

\bibitem{levine1966optimal}
W.~Levine and M.~Athans, ``On the optimal error regulation of a string of
  moving vehicles,'' \emph{IEEE Transactions on Automatic Control}, vol.~11,
  no.~3, pp. 355--361, 1966.

\bibitem{darbha1999}
S.~Darbha and K.~R. Rajagopal, ``Intelligent cruise control systems and traffic
  flow stability,'' \emph{Transportation Research Part C: Emerging
  Technologies}, vol.~7, no.~6, pp. 329 -- 352, 1999.

\bibitem{besselink2017string}
B.~Besselink and K.~H. Johansson, ``String stability and a delay-based spacing
  policy for vehicle platoons subject to disturbances,'' \emph{IEEE
  Transactions on Automatic Control}, 2017.

\bibitem{buehler2009darpa}
M.~Buehler, K.~Iagnemma, and S.~Singh, \emph{The DARPA urban challenge:
  autonomous vehicles in city traffic}.\hskip 1em plus 0.5em minus 0.4em\relax
  Springer, 2009, vol.~56.

\bibitem{shladover1995review}
S.~E. Shladover, ``Review of the state of development of advanced vehicle
  control systems ({AVCS}),'' \emph{Vehicle System Dynamics}, vol.~24, no. 6-7,
  pp. 551--595, 1995.

\bibitem{fenton1991automated}
R.~E. Fenton and R.~J. Mayhan, ``Automated highway studies at the {O}hio
  {S}tate {U}niversity-an overview,'' \emph{IEEE transactions on Vehicular
  Technology}, vol.~40, no.~1, pp. 100--113, 1991.

\bibitem{ioannou1993intelligent}
P.~Ioannou, Z.~Xu, S.~Eckert, D.~Clemons, and T.~Sieja, ``Intelligent cruise
  control: theory and experiment,'' in \emph{Proceedings of the 32nd IEEE
  Conference on Decision and Control}.\hskip 1em plus 0.5em minus 0.4em\relax
  IEEE, 1993, pp. 1885--1890.

\bibitem{rajamani1998design}
R.~Rajamani, S.~B. Choi, B.~K. Law, J.~K. Hedrick, R.~Prohaska, and P.~Kretz,
  ``Design and experimental implementation of control for a platoon of
  automated vehicles,'' \emph{{AMSE} Journal of Dynamic Systems, Measurement,
  and Control}, vol. 122, no.~3, pp. 470--476, 1998.

\bibitem{talebpour2016influence}
A.~Talebpour and H.~S. Mahmassani, ``Influence of connected and autonomous
  vehicles on traffic flow stability and throughput,'' \emph{Transportation
  Research Part C: Emerging Technologies}, vol.~71, pp. 143--163, 2016.

\bibitem{davis2004effect}
L.~C. Davis, ``Effect of adaptive cruise control systems on traffic flow,''
  \emph{Physical Review E}, vol.~69, no.~6, p. 066110, 2004.

\bibitem{xiao2011practical}
L.~Xiao and F.~Gao, ``Practical string stability of platoon of adaptive cruise
  control vehicles,'' \emph{IEEE Transactions on Intelligent Transportation
  Systems}, vol.~12, no.~4, pp. 1184--1194, 2011.

\bibitem{wang2018infrastructure}
M.~Wang, ``Infrastructure assisted adaptive driving to stabilise heterogeneous
  vehicle strings,'' \emph{Transportation Research Part C: Emerging
  Technologies}, vol.~91, pp. 276--295, 2018.

\bibitem{bose2001analysis}
A.~Bose and P.~Ioannou, ``Analysis of traffic flow with mixed manual and
  intelligent cruise control vehicles: Theory and experiments,'' California
  PATH, Tech. Rep. UCB-ITS-PRR-2001-13, 2001.

\bibitem{jin2018experimental}
I.~G. Jin, S.~S. Avedisov, C.~R. He, W.~B. Qin, M.~Sadeghpour, and G.~Orosz,
  ``Experimental validation of connected automated vehicle design among
  human-driven vehicles,'' \emph{Transportation Research Part C: Emerging
  Technologies}, vol.~91, pp. 335--352, 2018.

\bibitem{bansal2017forecasting}
P.~Bansal and K.~M. Kockelman, ``Forecasting {A}mericans' long-term adoption of
  connected and autonomous vehicle technologies,'' \emph{Transportation
  Research Part A: Policy and Practice}, vol.~95, pp. 49--63, 2017.

\bibitem{litman2017autonomous}
T.~Litman, \emph{Autonomous Vehicle Implementation Predictions}.\hskip 1em plus
  0.5em minus 0.4em\relax Victoria Transport Policy Institute Victoria, Canada,
  2017.

\bibitem{ioannou2005evaluation}
P.~A. Ioannou and M.~Stefanovic, ``Evaluation of acc vehicles in mixed traffic:
  Lane change effects and sensitivity analysis,'' \emph{IEEE Transactions on
  Intelligent Transportation Systems}, vol.~6, no.~1, pp. 79--89, 2005.

\bibitem{reuters}
\BIBentryALTinterwordspacing
{Business Insider}. (2018, August) These are the 20 best-selling cars and
  trucks in america in 2018. [Online]. Available:
  \url{https://www.businessinsider.com/best-selling-cars-and-trucks-in-america-in-2018-2018-8}
\BIBentrySTDinterwordspacing

\bibitem{bareket2003methodology}
Z.~Bareket, P.~S. Fancher, H.~Peng, K.~Lee, and C.~A. Assaf, ``Methodology for
  assessing adaptive cruise control behavior,'' \emph{IEEE Transactions on
  Intelligent Transportation Systems}, vol.~4, no.~3, pp. 123--131, 2003.

\bibitem{milanes2014modeling}
V.~Milan{\'e}s and S.~E. Shladover, ``Modeling cooperative and autonomous
  adaptive cruise control dynamic responses using experimental data,''
  \emph{Transportation Research Part C: Emerging Technologies}, vol.~48, pp.
  285--300, 2014.

\bibitem{oncu2014cooperative}
S.~Oncu, J.~Ploeg, N.~Van~de Wouw, and H.~Nijmeijer, ``Cooperative adaptive
  cruise control: Network-aware analysis of string stability,'' \emph{IEEE
  Transactions on Intelligent Transportation Systems}, vol.~15, no.~4, pp.
  1527--1537, 2014.

\bibitem{davis2013effects}
L.~C. Davis, ``The effects of mechanical response on the dynamics and string
  stability of a platoon of adaptive cruise control vehicles,'' \emph{Physica
  A: Statistical Mechanics and its Applications}, vol. 392, no.~17, pp.
  3798--3805, 2013.

\bibitem{liang1999optimal}
C.-Y. Liang and H.~Peng, ``Optimal adaptive cruise control with guaranteed
  string stability,'' \emph{Vehicle System Dynamics}, vol.~32, no. 4-5, pp.
  313--330, 1999.

\bibitem{milanes2014cooperative}
V.~Milan{\'e}s, S.~E. Shladover, J.~Spring, C.~Nowakowski, H.~Kawazoe, and
  M.~Nakamura, ``Cooperative adaptive cruise control in real traffic
  situations,'' \emph{IEEE Transactions on Intelligent Transportation Systems},
  vol.~15, no.~1, pp. 296--305, 2014.

\bibitem{seiler2004disturbance}
P.~Seiler, A.~Pant, and K.~Hedrick, ``Disturbance propagation in vehicle
  strings,'' \emph{IEEE Transactions on automatic control}, vol.~49, no.~10,
  pp. 1835--1842, 2004.

\bibitem{monteil2018l₂}
J.~Monteil, M.~Bouroche, and D.~J. Leith, ``$\mathcal{L}_2$ and
  $\mathcal{L}_\infty$ stability analysis of heterogeneous traffic with
  application to parameter optimization for the control of automated
  vehicles,'' \emph{IEEE Transactions on Control Systems Technology}, 2018.

\bibitem{shampine1997matlab}
L.~F. Shampine and M.~W. Reichelt, ``The {M}atlab {ODE} suite,'' \emph{SIAM
  journal on scientific computing}, vol.~18, no.~1, pp. 1--22, 1997.

\end{thebibliography}

\vspace{-0.5in}
\begin{IEEEbiography}[{\includegraphics[width=1in,height=1.25in,clip,keepaspectratio]{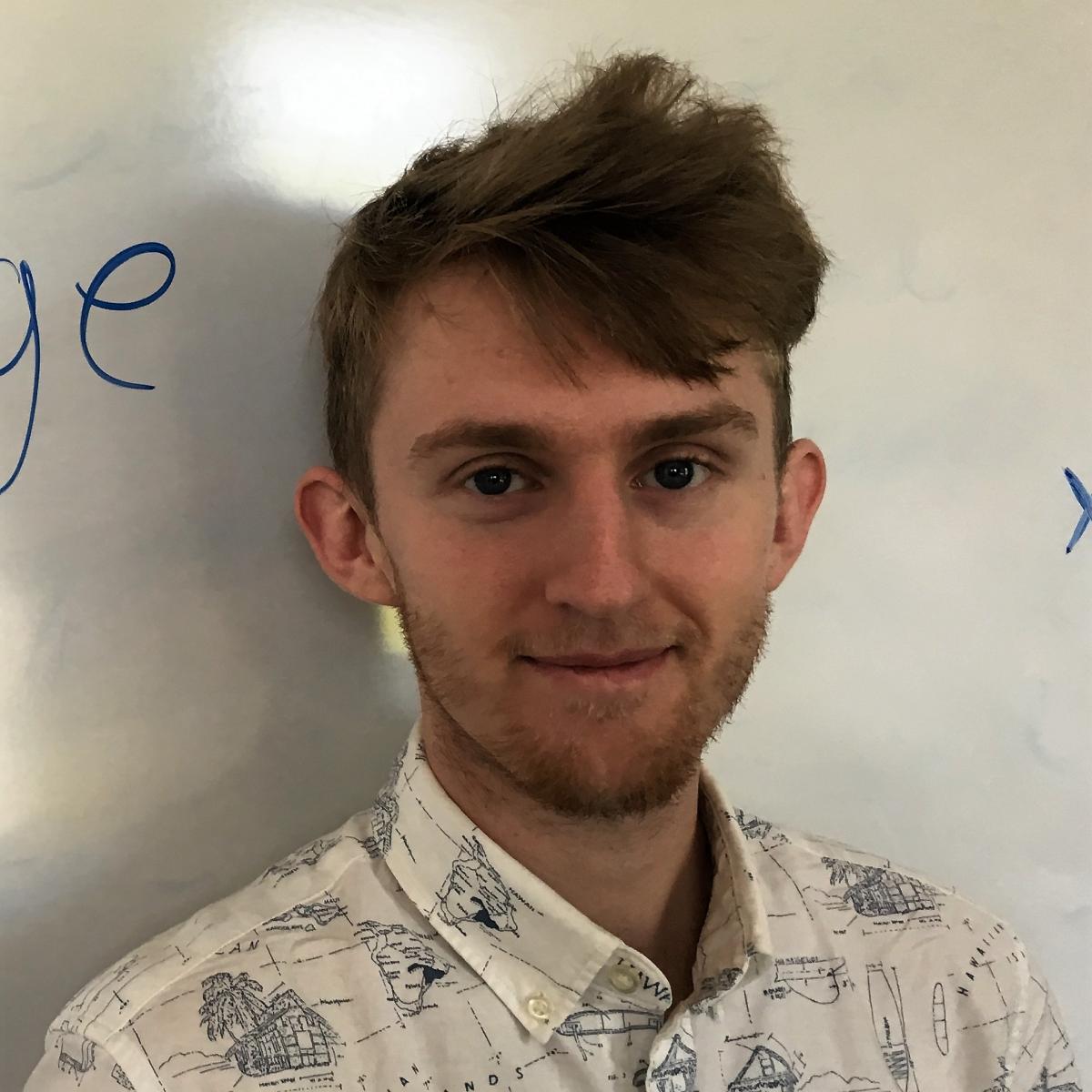}}]{George Gunter} is an undergraduate student in the Department of Civil and Environmental Engineering at the University of Illinois. George is also a visiting undergraduate researcher at the Institute for Software Integrated Systems at Vanderbilt University. His research interests include transportation cyber-physical systems and autonomous vehicles.
\end{IEEEbiography}

\vspace{-0.5in}

\begin{IEEEbiography}[{\includegraphics[width=1in,height=1.25in,clip,keepaspectratio]{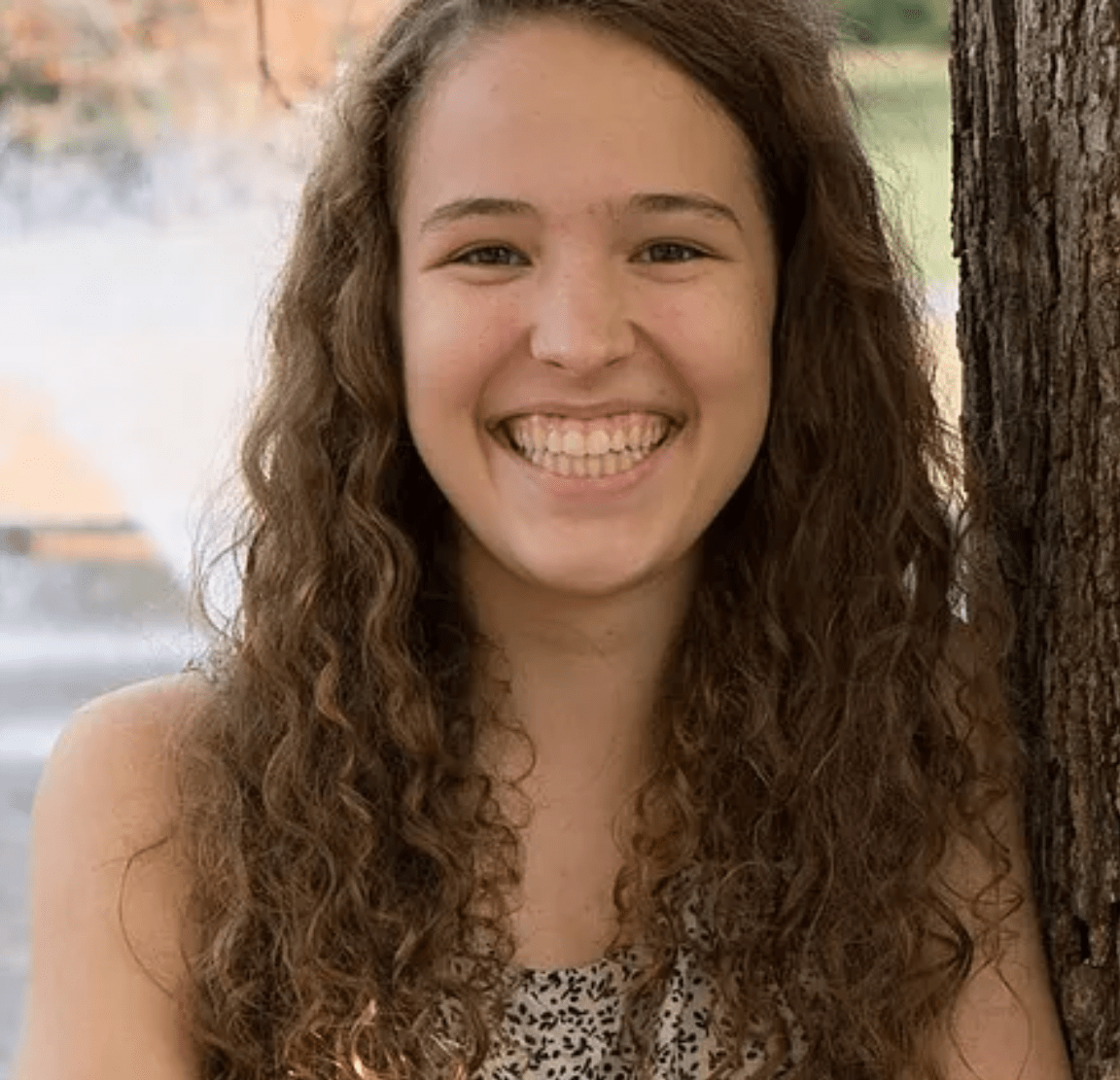}}]{Caroline Janssen} is an undergraduate student in the Department of Civil and Environmental Engineering at Vanderbilt University and an undergraduate researcher at the Institute for Software Integrated Systems. Caroline's research interests include transportation engineering, smart infrastructure, and autonomous vehicles. 
\end{IEEEbiography}

\vspace{-0.5in}

\begin{IEEEbiography}[{\includegraphics[width=1in,height=1.25in,clip,keepaspectratio]{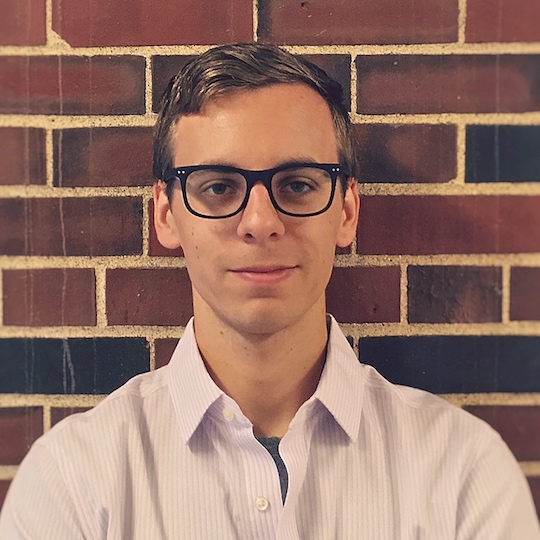}}]{William Barbour} is a Ph.D. student at Vanderbilt University and the Institute for Software Integrated Systems. William received a B.S. in Biosystems Engineering from the University of Tennessee, Knoxville, and a M.S. in Civil Engineering from the University of Illinois at Urbana-Champaign. William is a recipient of the Dwight David Eisenhower Transportation Fellowship from the Federal Highway Administration and the Eno Fellowship from the Eno Center for Transportation. His research interests include intelligent mobility systems, transportation sustainability, and smart cities. 
\end{IEEEbiography}
\vspace{-0.5in}

\begin{IEEEbiography}[{\includegraphics[width=1in,height=1.25in,clip,keepaspectratio]{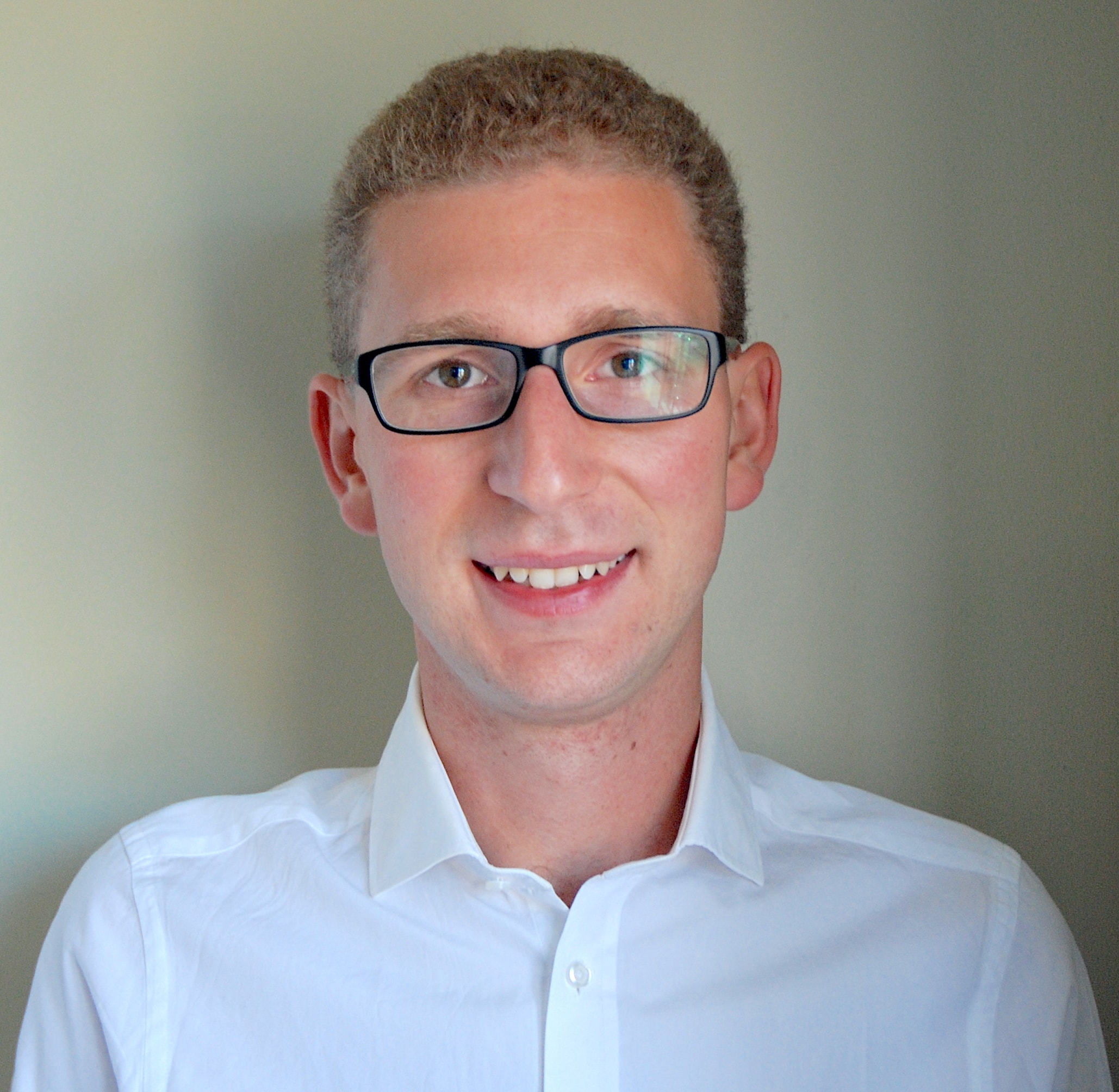}}]{Raphael Stern}
is a visiting researcher at the Institute for Software Integrated Systems at Vanderbilt University. Dr. Stern received a bachelor of science degree (2009), master of science degree (2015), and Ph.D. (2018) all in Civil Engineering from the University of Illinois at Urbana-Champaign. Dr. Stern has spent time as a visiting researcher at the Institute for Pure and Applied Mathematics at UCLA, and is a recipient of the Dwight David Eisenhower Graduate Fellowship from the Federal Highway Administration. Dr. Stern's research interests are in the area of traffic control and estimation with autonomous vehicles in the flow.
\end{IEEEbiography}
\vspace{-0.5in}

\begin{IEEEbiography}[{\includegraphics[width=1in,height=1.25in,clip,keepaspectratio]{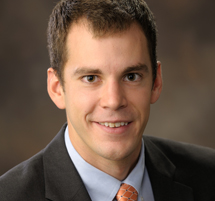}}]{Daniel B. Work}
is an associate professor in the Department of Civil and Environmental Engineering and the Institute for Software Integrated Systems at Vanderbilt University. Prof. Work earned his bachelor of science degree (2006) from the Ohio State University, and a master of science (2007) and Ph.D. (2010) from the University of California, Berkeley, each in civil engineering. Prof. Work has research interests in transportation cyber physical systems. Prof. Work's honors include being named a 2018 Gilbreth Lecturer by the the National Academy of Engineering, a 2014 CAREER award recipient from the National Science Foundation, and a recipient of the 2011 IEEE ITSS Best Dissertation Award.
\end{IEEEbiography}


\end{document}